\documentclass[trackchanges]{aastex701}

\newcommand{\CO}{\ensuremath{\rm ^{12}CO}}

\newcommand{\Kextv}{\ensuremath{K_V^{\rm ext}}}

\newcommand{\tw}{\ensuremath{T_{\rm w}}}
\newcommand{\mw}{\ensuremath{M_{\rm w}}}
\newcommand{\tc}{\ensuremath{T_{\rm c}}}
\newcommand{\mc}{\ensuremath{M_{\rm c}}}

\newcommand{\mum}{\ensuremath{\,{\rm \mu m}}}

\newcommand{\kpc}{\ensuremath{\,{\rm kpc}}}

\newcommand{\mpm}{\ensuremath{\,{\rm mag\, pc^{2}\, \msun^{-1}}}}

\newcommand{\jy}{\ensuremath{\,{\rm Jy}}}
\newcommand{\cg}{\ensuremath{\,{\rm cm^{2}\, g^{-1}}}}

\newcommand{\mcg}{\ensuremath{\,{\rm mag\, cm^{2}\, g^{-1}}}}
\newcommand{\msun}{\ensuremath{\,{M_\odot}}}
\newcommand{\kms}{\ensuremath{\,{\rm km\, s^{-1}}}}

\newcommand{\per}{\ensuremath{\,{\rm \%}}}
\usepackage{amsmath}
\usepackage{booktabs}
\usepackage{tablefootnote}
\usepackage{subfigure}
\usepackage{graphicx}
\usepackage[utf8]{inputenc}
\usepackage{CJK}
\hypersetup{colorlinks=true}
\defcitealias{Zhang2026AJ}{I}
\usepackage{soul}
\usepackage[normalem]{ulem}
\usepackage{bm}
\usepackage{multirow}

\begin{document}

\title{Estimation of Dust Mass from Infrared Emission and Extinction of Supernova Remnants: G93.7$-$0.2, G109.1$-$1.0, G156.2+5.7, and G166.0+4.3}


\begin{CJK*}{UTF8}{gbsn}
\correspondingauthor{Jun Li, Biwei Jiang}
\email{lijun@gzhu.edu.cn, bjiang@bnu.edu.cn}

\author[0000-0002-1007-3700]{Zhe Zhang (张哲)}
\affiliation{School of Physics and Astronomy, Beijing Normal University, Beijing 100875, People's Republic of China}
\affiliation{Institute for Frontiers in Astronomy and Astrophysics, Beijing Normal University, Beijing 102206, People's Republic of China}
\email{zhangz324@mail.bnu.edu.cn}

\author[0000-0001-9328-4302]{Jun Li (李军)}
\affiliation{Center for Astrophysics, Guangzhou University, Guangzhou 510006, People's Republic of China}
\affiliation{National Astronomical Observatories, Chinese Academy of Sciences, Beijing 100101, People's Republic of China}
\email{lijun@gzhu.edu.cn}

\author[0000-0003-3168-2617]{Biwei Jiang (姜碧沩)}
\affiliation{School of Physics and Astronomy, Beijing Normal University, Beijing 100875, People's Republic of China}
\affiliation{Institute for Frontiers in Astronomy and Astrophysics, Beijing Normal University, Beijing 102206, People's Republic of China}
\email{bjiang@bnu.edu.cn}

\author[0000-0003-2645-6869]{He Zhao (赵赫)}
\affiliation{Institute of Astronomy and Physics, Inner Mongolia University, Hohhot 010021, People's Republic of China}
\affiliation{Departamento de Fisica y Astronomia, Facultad de Ciencias Exactas, Universidad Andres Bello, Fernandez Concha 700, 8320000 Santiago, Chile}
\email{he.zhao@oca.eu}


\begin{abstract}

Supernova remnants (SNRs) are major sites for both the production and destruction of interstellar dust, and quantifying their dust budget is essential for understanding the life cycle of cosmic dust. In this work, the dust masses of four Galactic SNRs (G93.7$-$0.2, G109.1$-$1.0, G156.2+5.7, and G166.0+4.3) are estimated using two complementary methods: the three-dimensional (3D) interstellar extinction map and infrared (IR) spectral energy distribution (SED) fitting based on photometry from WISE, IRAS, AKARI, and Planck.
The extinction masses, derived from the differential extinction within each SNR's distance interval, are 108.3, 82.0, 48.8, and 119.2$\msun$, respectively. A two-component (``warm + cold") modified blackbody fitting yields warm dust temperatures of 43--74\,K and cold dust temperatures of 13--16\,K, with the cold dust component dominating the total IR-emission mass ($\sim$90--400$\msun$). The extinction masses and IR emission masses show systematic differences, likely caused by sightline contamination from unrelated foreground/background material and uncertainties in dust temperatures and opacities.

\end{abstract}

\keywords{Supernova remnants(1667); Interstellar dust(836); Interstellar extinction(841); Dust continuum emission(412)}


\section{Introduction} 
Supernovae (SNe) are the major sources of heavy elements and dust to the interstellar medium (ISM), as newly formed dust grains can condense in the cooling ejecta \citep{Gomez2012ApJ,deLooze2017MNRAS}. Observations and theoretical models indicate that each SN can produce approximately 0.1--1$\msun$ of ejecta dust \citep{Barlow2010AA,Matsuura2011Sci}. However, the strong forward and reverse shocks generated by SN explosions also efficiently destroy dust via thermal sputtering and grain--grain collisions \citep{Jones1994ApJ,Dwek1996ApJ,Slavin2015ApJ}. Small dust grains are particularly fragile, while sufficiently large grains may survive and be injected into the ISM, thereby reshaping the grain size distribution \citep{Nozawa2007ApJ,Zhao2025ApJL}. In dense clumps with efficient cooling, dust-assisted cooling can further promote the survival of dust grains \citep{Gonzalez2025AA}. More than 90$\per$ of the surviving dust mass resides in cool shocked gas, and only large grains ($\sim$ 0.1$\mum$) survive in the hot shocked ISM \citep{Priestley2022MNRAS}. Since evolved supernova remnants (SNRs) are often dominated by swept-up interstellar dust rather than fresh ejecta dust, accurately measuring dust masses in SNRs is crucial for quantifying dust destruction by SN shocks and understanding the life cycle of interstellar matter in galaxies.

Infrared (IR) observations provide one of the most powerful methods for studying dust in SNRs, because the IR continuum emission is primarily produced by thermal radiation from dust grains \citep{Dwek1987ApJ,Chawner2019MNRAS}. The most widely used method of estimating dust masses is fitting the IR spectral energy distribution (SED) of dust emission, which has been applied to Galactic SNRs \citep{Gomez2012MNRAS,Temim2017ApJ,Priestley2022MNRAS}, SNRs in the Large Magellanic Cloud \citep{Lakicevic2015ApJ}, and SNR candidates in M31 \citep{Wang2022AJ}. Nevertheless, this method suffers from significant uncertainties. First, the derived dust mass strongly depends on the assumed dust temperature, which is often oversimplified as a constant. Second, it relies on the adopted dust emissivity law, which is model-dependent. Third, IR emission cannot distinguish dust intrinsic to the SNR from that in foreground or background clouds along the sightline, leading to potential overestimation \citep{Li2022ApJ}.

An alternative approach is based on interstellar extinction, which measures dust column density via absorption and scattering of background starlight \citep{Zhao2018ApJ,Wang2020AA}. Extinction is largely independent of dust temperature and can even constrain the dust emissivity through extinction curve model. Extinction mass thus provides a complementary and cross-checking method. However, extinction also suffers from limitations: it cannot easily distinguish dust associated with the SNR from dust residing in physically unrelated molecular clouds (MCs) along the same sightline, and high spatial resolution requires a sufficiently dense stellar background, which is often lacking in sparse regions \citep{Chen2017MNRAS,Zhao2020ApJ}.

Given the complementary strengths and weaknesses of IR emission and extinction methods, combining them offers more robust constraints on dust masses. Recently, \citet{Li2022ApJ} applied this combined approach to the SNR IC 443, deriving dust masses of $\sim$
46$\msun$ from IR emission and $\sim$
66$\msun$ from extinction. Their work demonstrated how the two methods can validate each other and reduce systematic uncertainties.

Reliable distance measurements are essential for converting observed fluxes and extinctions into accurate dust masses. In a companion paper \citep{Zhang2026AJ} (hereafter Paper \citetalias{Zhang2026AJ}), we studied four Galactic SNRs (G93.7$-$0.2, G109.1$-$1.0, G156.2+5.7, and G166.0+4.3) and determined their distances by identifying physically associated MCs. Specifically, Paper \citetalias{Zhang2026AJ} combined CO line observations with three-dimensional (3D) extinction maps, and used the morphological similarity between CO integrated intensity maps and extinction bin maps to identify interacting MCs. This method reliably constrained the distances to these four SNRs, providing a solid foundation for accurate dust mass determinations.

In this work, we build upon the distance results of Paper \citetalias{Zhang2026AJ} to derive dust masses for the same four SNRs using both extinction measurements and IR SED fitting. We first estimate the dust masses of each SNR within the selected distance interval from the 3D extinction map, which directly traces dust column density and is independent of dust temperature. We then determine the dust temperatures and masses from IR emission by fitting a ``warm + cold" dust model, and further examine the spatial distributions of the two components on a pixel-by-pixel basis. The masses obtained from the two methods are compared in order to assess possible contamination from foreground or background clouds along the sightline.

This paper is organized as follows. In Section \ref{sec:2data}, the datasets and reduction are described.
Section \ref{extinction result} presents the SNR dust masses based on the extinction map. Section~\ref{sec: infrared} reports dust temperatures and masses from IR SED fitting. Discussion follows in Section~\ref{sec:discuss}, and conclusions are given in Section~\ref{sec:sum}.

\section{Data and Reduction} \label{sec:2data}

\subsection{Target SNRs} \label{sec2:sample}
The targets of this study are the four Galactic SNRs G93.7$-$0.2, G109.1$-$1.0, G156.2+5.7, and G166.0+4.3, whose basic parameters are summarized in Table~\ref{table1}. 
G93.7$-$0.2 has been suggested to have already entered the radiative phase \citep{Uyaniker2002ApJ}. The evolutionary status of G109.1$-$1.0, on the other hand, remains under debate. Although the remnant is generally regarded as being in the adiabatic (Sedov--Taylor) phase, \citet{Sanchez2018MNRAS} reported prominent optical emission from relatively cool ionized gas, indicating that localized radiative cooling has already begun within dense regions of the remnant. G156.2+5.7 is a typical remnant evolving in a low-density environment and is currently in the Sedov phase \citep{Uchida2012PASJ}. G166.0+4.3 appears to be transitioning from the Sedov to the radiative phase, indicative of a relatively evolved stage \citep{Arias2019AA, Xiao2023ApJ}.
These remnants represent a range of environmental conditions, including interactions with MCs, expansion into low-density cavities, and evolution within structured interstellar environments. Such diversity provides an opportunity to investigate how dust properties vary among SNRs under different ambient conditions.

\subsection{Infrared Data} \label{sec2:ir_data}
To perform spectral energy distribution (SED) fitting over a broad wavelength range, we make use of archival photometric data from four space-based facilities: the Wide-field Infrared Survey Explorer (WISE), the Infrared Astronomical Satellite (IRAS), the AKARI Far-Infrared All-Sky Survey, and the Planck all-sky survey.
WISE provides photometry in four bands centered at 3.4, 4.6, 12, and 22$\mum$ \citep{Wright2010AJ}, of which only the 12 and 22$\mum$ bands are adopted in this work, since shorter wavelengths are dominated by stellar emission. IRAS covers 12, 25, 60, and 100$\mum$ \citep{Neugebauer1984ApJ}, and AKARI covers 65, 90, 140, and 160$\mum$ \citep{Doi2015PASJ}. From the Planck mission, we adopt the four highest-frequency bands at 350, 550, 850, and 1380$\mum$ \citep{Planck2016AA}, which are most sensitive to thermal emission from cold dust. The angular resolutions, absolute calibration uncertainties, and original pixel sizes of all bands are listed in Table~\ref{table2}. Together, these data sets span from the mid-IR to the millimeter regime. All images used in this work are retrieved from the NASA/IPAC Infrared Science Archive (IRSA)\footnote{\url{https://irsa.ipac.caltech.edu/frontpage/}}.

\subsection{Three-Dimensional Extinction Map} \label{sec2:3D Exti}
To estimate the dust mass associated with each SNR independently of its infrared emission, we employ the 3D extinction map of \citet{Green2019ApJ}, which is constructed from stellar photometry and parallaxes drawn from Gaia, Pan-STARRS~1, and 2MASS. The map covers the sky regions of all four SNRs and provides a relatively high angular resolution of 3.4\arcmin\ per voxel. Its maximum reliable distance ranges from several hundred parsecs to more than 10$\kpc$, depending on the stellar density along each sightline. We query the map using the \texttt{BayestarQuery} class provided in the Python package \texttt{dustmaps}\footnote{\url{https://dustmaps.readthedocs.io/en/latest/}} \citep{Green2018JOSS}.


\section{Dust Mass from Extinction} \label{extinction result}

\subsection{Method}\label{sec:ext_method}

Using the 3D dust extinction map of \citet{Green2019ApJ}, Paper~\citetalias{Zhang2026AJ} constructed extinction--distance profiles toward the four SNRs (G93.7$-$0.2, G109.1$-$1.0, G156.2+5.7, and G166.0+4.3). Significant extinction jumps along each sightline were identified, and the SNR distances were then determined by identifying the distance bins where the differential extinction structures show the strongest spatial correspondence with the CO velocity-integrated intensity maps. Since the extinction enhancement produced by MCs is generally much stronger than that of the diffuse background ISM, these extinction jumps provide reliable kinematic-independent distance constraints to the interacting SNR-MC systems.

In addition to providing distance estimates, the 3D extinction map also traces the dust column density along each sightline. Since the interstellar extinction is proportional to the dust column density, the differential extinction within the SNR distance interval can be converted into a dust mass surface density and, in turn, used to estimate the total dust mass of the SNR distance interval. Following \citet{Zhao2018ApJ}, we compute the extinction-based dust mass as
\begin{equation}\label{equ:mass_ext}
	M_{\rm d}^{\rm ext} = \Sigma_{\rm dust}\, A_{\rm SNR}
	= \frac{A_V}{\Kextv}\, A_{\rm SNR}
\end{equation}
where $\Sigma_{\rm dust}$ is the dust mass surface density, $A_V$ is the differential $V$-band extinction associated with the SNR distance interval, $\Kextv$ is the dust mass extinction coefficient in the $V$ band (in units of $\mcg$), $A_{\rm SNR} = \pi R_{\rm SNR}^{2}$ is the projected area of the remnant.

The physical radius (or semi-major axis) of each SNR, $R_{\rm SNR}$, is derived from its distance $D$ and angular aperture radius (defined in Section~\ref{IR}) as $R_{\rm SNR} \approx D\,\theta$, where $\theta = R_{\rm aperture}$ in radians; the resulting values are listed in Table~\ref{table1}. 

The extinction coefficient $\Kextv$ depends on the composition and size distribution of dust grains. Adopting a graphite--silicate grain mixture with a size distribution constrained to reproduce the observed Galactic extinction curve, \citet{Nozawa2013ApJ} derived $\Kextv = (3.7 \pm 0.5) \times 10^{4}\,\mcg$, equivalent to $7.75 \pm 1.05\,\mpm$. We adopt this value throughout the present work.

\subsection{Results}\label{sec:ext_results}

Based on the extinction analysis of Paper~\citetalias{Zhang2026AJ}, the distance and corresponding extinction interval for each SNR are stated below: $D = 1.82\,\kpc$ and $1.66$--$1.98\,\kpc$ for G93.7$-$0.2; $3.05\,\kpc$ and $2.86$--$3.25\,\kpc$ for G109.1$-$1.0; $0.60\,\kpc$ and $0.40$--$0.79\,\kpc$ for G156.2+5.7; and $3.44\,\kpc$ and $3.16$--$3.73\,\kpc$ for G166.0+4.3. To minimize the influence of the noisier edges of the extinction--distance profiles, we adopt the central 80\% of each distance interval when computing the extinction mass. The differential extinction within this restricted interval is then taken to represent the dust column directly associated with each SNR and its surrounding MCs.

\begin{figure}[ht!]
\centering
\includegraphics[width=1.0\textwidth]{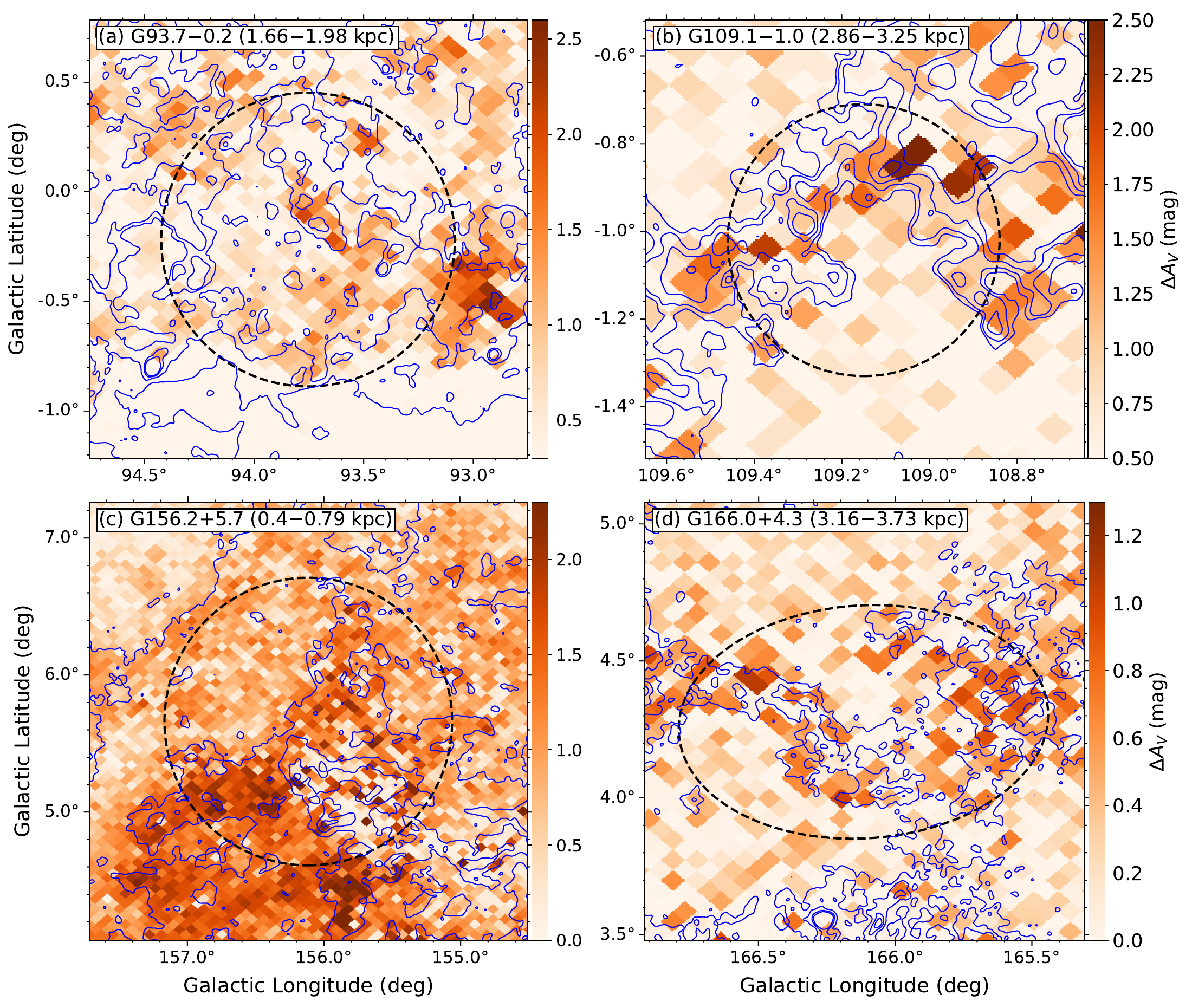}
\caption{Differential extinction maps in the distance bins associated with the MCs interacting with each SNR, based on the distance slicing of Paper~\citetalias{Zhang2026AJ}. The black dashed circle marks the photometric aperture adopted in this work. Blue contours show the \textit{AKARI} 140\,$\mu$m emission, with levels of 120, 150, 180, and 210\,MJy\,sr$^{-1}$ for G93.7$-$0.2; 170, 180, 200, and 230\,MJy\,sr$^{-1}$ for G109.1$-$1.0; 30, 45, 60, 75, and 90\,MJy\,sr$^{-1}$ for G156.2+5.7; and 32, 38, 44, and 50\,MJy\,sr$^{-1}$ for G166.0+4.3. The visual extinction \(A_{V}\) is obtained from the reddening map of \citet{Green2019ApJ} via \(A_{V}=R_{V}\,E(B-V)\) with \(R_{V}=2.742\) \citep{Schlafly2011ApJ}.
}
\label{fig1}
\end{figure}

The resulting extinction maps are shown in Figure~\ref{fig1}. The blue contours trace the AKARI 140\,$\mum$ emission, and the black dashed circles mark the SNR apertures defined in Section~\ref{IR}. A clear spatial correspondence is seen between the far-IR emission and the extinction structures across all four remnants.
Although there are foreground/background MCs at the sightline of the SNRs, as revealed by the analysis of 3D extinction in Paper \citetalias{Zhang2026AJ}, this morphological consistency suggests that the 140$\mum$ emission mainly comes from the dust clouds physically associated with the SNRs. 

Integrating the extinction-derived dust column density within each SNR aperture following Equation~\ref{equ:mass_ext}, we obtain dust masses of $M_{\rm d}^{\rm ext} = 108.3^{+17.9}_{-13.5}\,\msun$ for G93.7$-$0.2, 
$82.0^{+13.6}_{-10.2}\,\msun$ for G109.1$-$1.0, 
$48.8^{+8.1}_{-6.1}\,\msun$ for G156.2+5.7, 
and $119.2^{+19.7}_{-14.8}\,\msun$ for G166.0+4.3. 
The quoted uncertainties are propagated solely from the formal uncertainty in $\Kextv$ adopted in Section~\ref{sec:ext_method}. However, the extinction masses are also subject to additional systematic effects, including the choice of distance interval over which the differential extinction is integrated, intrinsic uncertainties of the Bayestar19 3D extinction map \citep{Green2019ApJ}, the adopted value of $R_V$, the definition of the SNR aperture, and the uncertainty in the SNR distance itself. A detailed quantification of these systematic effects is beyond the scope of the present study, and they are therefore not folded into the quoted error bars.

\section{Dust Mass from Infrared Emission} \label{sec: infrared}

\subsection{Infrared Photometry} \label{IR}

The integrated infrared fluxes of the four SNRs are measured by aperture photometry using the \texttt{Photutils} Python package \citep{Bradley2024zndo}\footnote{\url{https://photutils.readthedocs.io/en/stable/}}. For each remnant, the source aperture and background annulus are shown in every band as yellow solid and dashed circles respectively, with the central coordinates adopted from \citet{Green2025JApA}. The aperture and annulus radii are listed in Table~\ref{table3}, and the source aperture sizes are determined from the radio continuum images compiled in the SNRcat database \citep{Ferrand2012AdSpR}. The integrated flux density of each SNR is measured within its source aperture, and the background, estimated from the surrounding annulus, is subtracted. Following \citet{Li2022ApJ}, we do not include aperture and color corrections in our error budget, as they are negligible compared with the dominant photometric uncertainties. The resulting flux densities for all four SNRs are summarized in Table~\ref{table2}.

\begin{figure}[h!]
\centering
\includegraphics[width=0.9\textwidth]{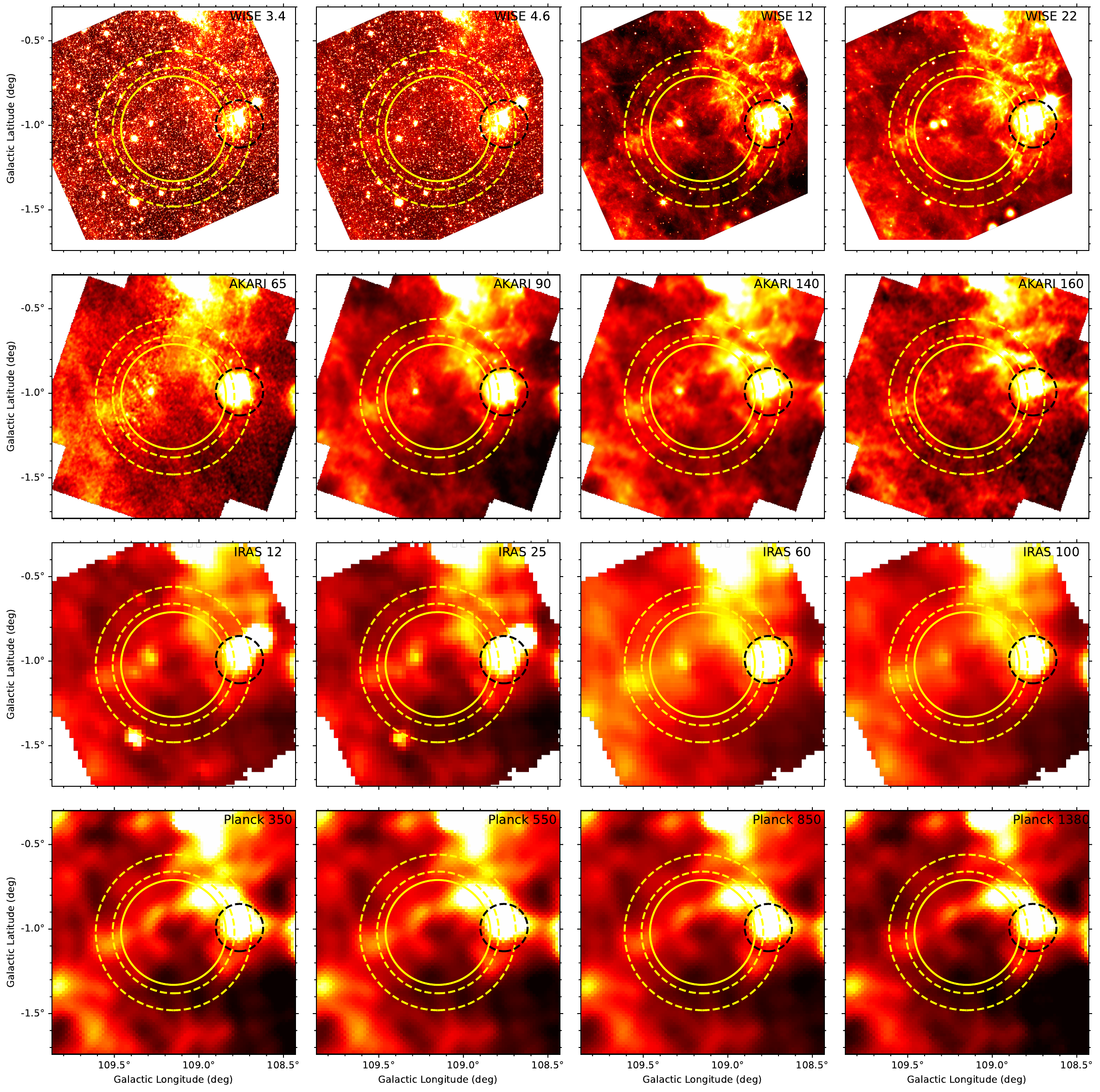}
\caption{Multi-band images of G109.1$-$1.0 from \textit{WISE} (3.4, 4.6, 12, and 22\,$\mu$m), \textit{AKARI} (65, 90, 140, and 160\,$\mu$m), \textit{IRAS} (12, 25, 60, and 100\,$\mu$m), and \textit{Planck} (350, 550, 850, and 1380\,$\mu$m). The yellow solid circle shows the photometric aperture, centred at \((l,b)=(109\fdg15,\,-1\fdg02)\) with a radius of \(18\farcm6\); the yellow dashed circles mark the sky-background annulus, with inner and outer radii of \(21\farcm6\) and \(27\farcm6\) respectively. The black dashed circle indicates the H\,{\sc ii} region S152, which is masked out before photometry is performed. Each image is shown at its original angular resolution and pixel scale.}
\label{fig2}
\end{figure}

To minimize the influence of localized bright structures within the background annuli, the median (rather than the mean) of the pixel values is adopted as the background level. This treatment is particularly important for G109.1$-$1.0, whose background annulus partially encompasses a bright complex to the southwest that is detected in all IR bands. This region hosts active star formation, young stellar objects, and the \mbox{H\,{\sc ii}} region S152 \citep{Sharpless1959ApJS,Gregory1980Nature,Kahane1985AA}, and would otherwise bias the background estimate upward. We therefore mask this region during the photometric measurements, as indicated by the black dashed circle in Figure~\ref{fig2}.

Bands shortward of $\sim$20\,$\mum$ are excluded from the subsequent SED fitting because they are strongly contaminated by ionic and molecular line emission as well as by stellar continuum \citep{Li2022ApJ}. At wavelengths $\gtrsim 20\,\mum$, line emission no longer dominates the broadband flux, and we therefore apply no line-emission correction. However, toward the submillimeter regime, the synchrotron component becomes increasingly important, although it is negligible at shorter IR wavelengths. To remove this contribution, we model the radio synchrotron emission of each SNR as a single power law, $S_\nu \propto \nu^{-\alpha}$, with the spectral index $\alpha$ and the 1\,GHz flux density $S_{1\,\rm GHz}$ taken from \citet{Green2025JApA}: $(\alpha,\,S_{1\,\rm GHz}) = (0.65,\,65\,\jy)$ for G93.7$-$0.2, $(0.45,\,20\,\jy)$ for G109.1$-$1.0, $(0.50,\,5\,\jy)$ for G156.2+5.7, and $(0.37,\,7\,\jy)$ for G166.0+4.3. The fitted radio power law is then extrapolated to the mid-IR and submillimeter bands and subtracted from the measured fluxes prior to SED fitting.

\subsection{SED Modeling} \label{sec:sed_model}

After subtracting the synchrotron contribution (Section~\ref{IR}), the IR SED of each SNR is fitted with a two-component modified blackbody model following \citet{Li2022ApJ}:
\begin{equation}\label{equ:sed_fit}
	F_\nu(\lambda) = \frac{M_{\rm w}}{D^2}\,\kappa(a,\lambda)\,B_\nu(\lambda,T_{\rm w})
	+ \frac{M_{\rm c}}{D^2}\,\kappa(a,\lambda)\,B_\nu(\lambda,T_{\rm c}),
\end{equation}
where \(M_{\rm w}\) and \(T_{\rm w}\) (\(M_{\rm c}\) and \(T_{\rm c}\)) denote the mass and temperature of the warm (cold) dust component, \(B_\nu(\lambda,T)\) is the Planck function, and \(D\) is the distance to the remnant. The mass absorption coefficient \(\kappa(a,\lambda)\) depends on the dust grain radius \(a\) through
\begin{equation}\label{equ:kapp}
	\kappa(a,\lambda) = \frac{3\,Q(a,\lambda)}{4\rho a},
\end{equation}
where \(Q(a,\lambda)\) is the absorption efficiency and \(\rho\) is the mass density of the grain material. For grains smaller than \(1\,\mum\), the values of \(\kappa\) for silicate and graphite in the mid-IR and far-IR are essentially independent of \(a\) \citep{Hildebrand1983QJRAS}, and we therefore fix \(a = 0.1\,\mum\) in Equation~\ref{equ:sed_fit}.

The mass absorption coefficient can equivalently be expressed as a power law,
\begin{equation}\label{equ:kapp2}
	\kappa(\lambda) = \kappa_{\lambda_0}\left(\frac{\lambda_0}{\lambda}\right)^{\beta},
\end{equation}
where \(\kappa_{\lambda_0}\) is the reference value at wavelength \(\lambda_0\) and \(\beta\) is the dust emissivity spectral index. We adopt \(\lambda_0 = 500\,\mum\) because of its widespread use and its weak sensitivity to dust temperature variations \citep{Gomez2012MNRAS,Clark2016MNRAS}. The dust grains are modeled as a mixture of silicate and graphite with a fixed mass ratio \(M_{\rm sil}/M_{\rm gra}=2\) \citep{Draine1984ApJ}, corresponding to \(\kappa_{500}=1.45\,\cg\) and \(\beta=2.04\) \citep{Li2022ApJ}. The Planck \(1380\,\mum\) band is excluded from the fit because it lies beyond the wavelength range over which the dust optical constants are defined.

\subsubsection{Integrated SED Fitting Results} \label{sec:sed_results}

\begin{figure}[ht!]
\centering
\includegraphics[width=1.0\textwidth]{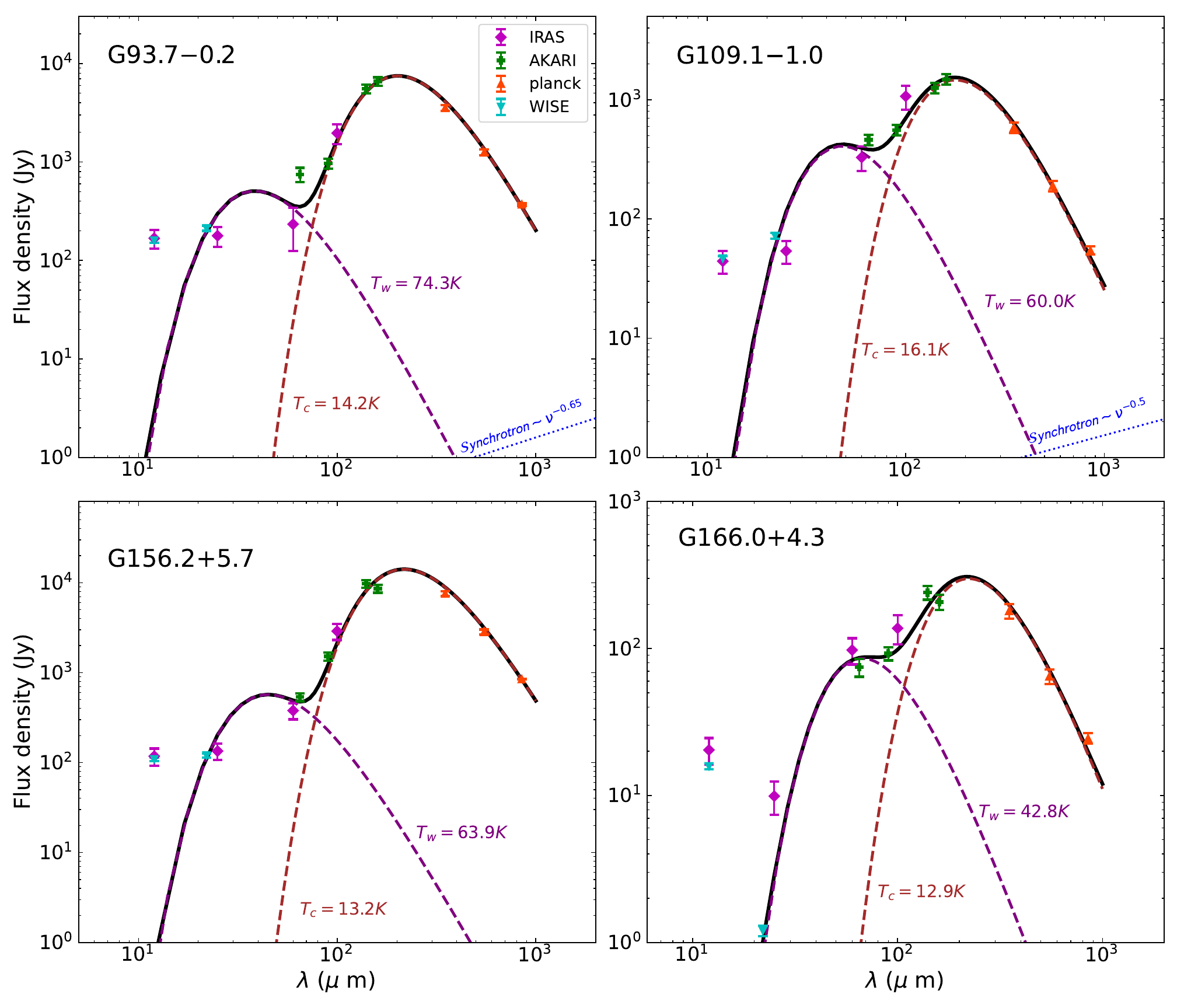}  
 \caption{The SEDs of the four SNRs from the near-IR to the submillimeter, together with the best-fit models. Filled points with the error bars (calibration uncertainties) show the photometric measurements without subtracting the contribution from line emission. A two-component modified-blackbody model is fitted to the data at $\lambda>20\,\mu$m in each panel: the warm and cold components are shown as purple and brown dashed lines, respectively, and the violet dotted line denotes the synchrotron contribution. The black solid line is the sum of the two dust components and the synchrotron emission.}
\label{fig3}
\end{figure}

The best-fitting IR SEDs for the four remnants are shown in Figure~\ref{fig3}. In every case, the observed emission is well reproduced by the sum of a warm and a cold component, yielding the characteristic double-peaked IR SED. The warm component dominates the mid-IR, whereas the cold component takes over at far-IR and submillimeter wavelengths.

The uncertainties in the fitted dust masses and temperatures are estimated through a bootstrap procedure: for each remnant, the observed fluxes are resampled within their Gaussian uncertainties and refitted, with \(2000\) realizations performed in total. The reported errors correspond to the \(68\%\) confidence intervals of the resulting distributions. The derived temperatures and masses are summarized in Table~\ref{table4}. For all four remnants, the cold dust mass exceeds the warm dust mass by two to three orders of magnitude, indicating that the total dust mass is dominated by the cold component. The warm dust contributes only a small fraction of the total IR-emission mass, and is especially low in G93.7\(-\)0.2 and G156.2+5.7.

The two-temperature fits yield warm and cold dust components with substantially different temperatures and masses. The warm component should be interpreted with caution, because small grains in shocked plasmas can undergo stochastic heating and experience transient temperature excursions, producing strong mid-IR emission that cannot be described by a single equilibrium temperature \citep{Purcell1976ApJ,Dwek1986ApJ,Guhathakurta1989ApJ}. \citet{Priestley2019MNRAS} used DINAMO to fit the observed Cas A SED with radiative and collisional grain heating.
Therefore, the warm dust temperature and mass derived from the two-temperature model should be regarded as approximate and model-dependent parameters rather than accurate calculation. 
Nevertheless, because the total dust mass is overwhelmingly dominated by the cold component, the uncertainty associated with stochastic heating is unlikely to significantly affect the total dust mass estimate \citep{Chiang2018ApJ}. In addition, to avoid contamination from line emission at wavelengths shortward of 20$\mum$, we do not include very hot dust in our SED fitting in this work.

Because these remnants are partially embedded in molecular environments, the integrated IR SED cannot unambiguously distinguish dust physically associated with the SNR from unrelated foreground/background material. To better separate these components, we therefore turn to a spatially resolved analysis of the dust properties.

\subsubsection{Spatially Resolved Dust Properties} \label{sec:sed_maps}

To investigate the spatial variation of the dust properties, we perform pixel-by-pixel SED fitting using the same two-component modified blackbody model defined in Equation~\ref{equ:sed_fit}. Prior to fitting, all background-subtracted images are convolved with a Gaussian kernel to match the angular resolution of the IRAS \(100\,\mum\) band (\(300\arcsec\)) and are regridded to a common pixel scale of \(60\arcsec\). Only pixels with positive flux densities in all bands are retained, in order to ensure reliable SED constraints.

The flux uncertainties of individual pixels are assumed to be dominated by absolute calibration errors. Following \citet{Li2022ApJ}, we adopt relative uncertainties of \(20\%\) for the IRAS bands, \(10\%\) for the AKARI bands, \(5.6\%\) for the WISE \(22\,\mum\) band, and \(8.2\%\), \(7.6\%\), and \(4.7\%\) for the Planck \(350\), \(550\), and \(850\,\mum\) bands, respectively.

\begin{figure}[ht!]
\centering
\includegraphics[width=1.0\textwidth]{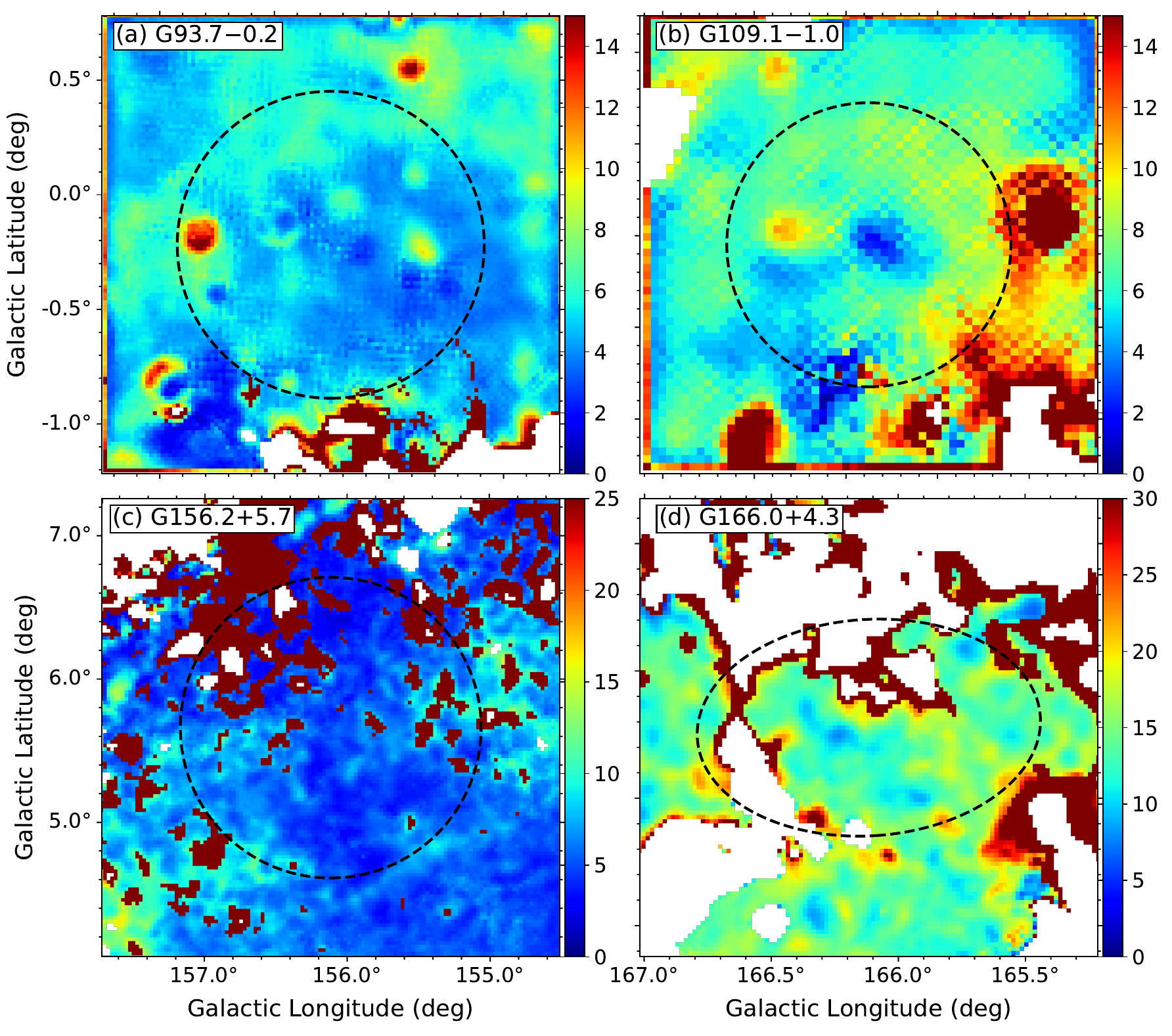}
\caption{The $\chi^2$/dof distribution derived from fitting the MIR to submillimeter SEDs of four SNRs pixel-by-pixel. The dashed regions indicate the photometry apertures for all SNRs.}
\label{fig4}
\end{figure}

The goodness of fit at each pixel is quantified by the reduced chi-square,
\begin{equation}
	\frac{\chi^2}{\rm dof}
	=\frac{1}{N_{\rm obs}-N_{\rm par}}
	\sum_{j=1}^{N_{\rm obs}}
	\frac{\left[F_\nu^{\rm mod}(\lambda_j)-F_\nu^{\rm obs}(\lambda_j)\right]^2}
	{\sigma^2(\lambda_j)},
\end{equation}
where \(F_\nu^{\rm mod}(\lambda_j)\) and \(F_\nu^{\rm obs}(\lambda_j)\) are the model and observed pixel flux densities, \(\sigma(\lambda_j)\) 
is the corresponding flux uncertainty, and \(N_{\rm obs}=11\) and \(N_{\rm par}=4\) are the numbers of data points and free parameters, respectively. The resulting \(\chi^2/{\rm dof}\) maps are presented in Figure~\ref{fig4}, and pixels with elevated \(\chi^2/{\rm dof}\) are mainly located near the map edges and are attributable to imperfect background subtraction and to a low signal-to-noise ratio.

\begin{figure}[ht!]
\centering
\includegraphics[width=1.0\textwidth]{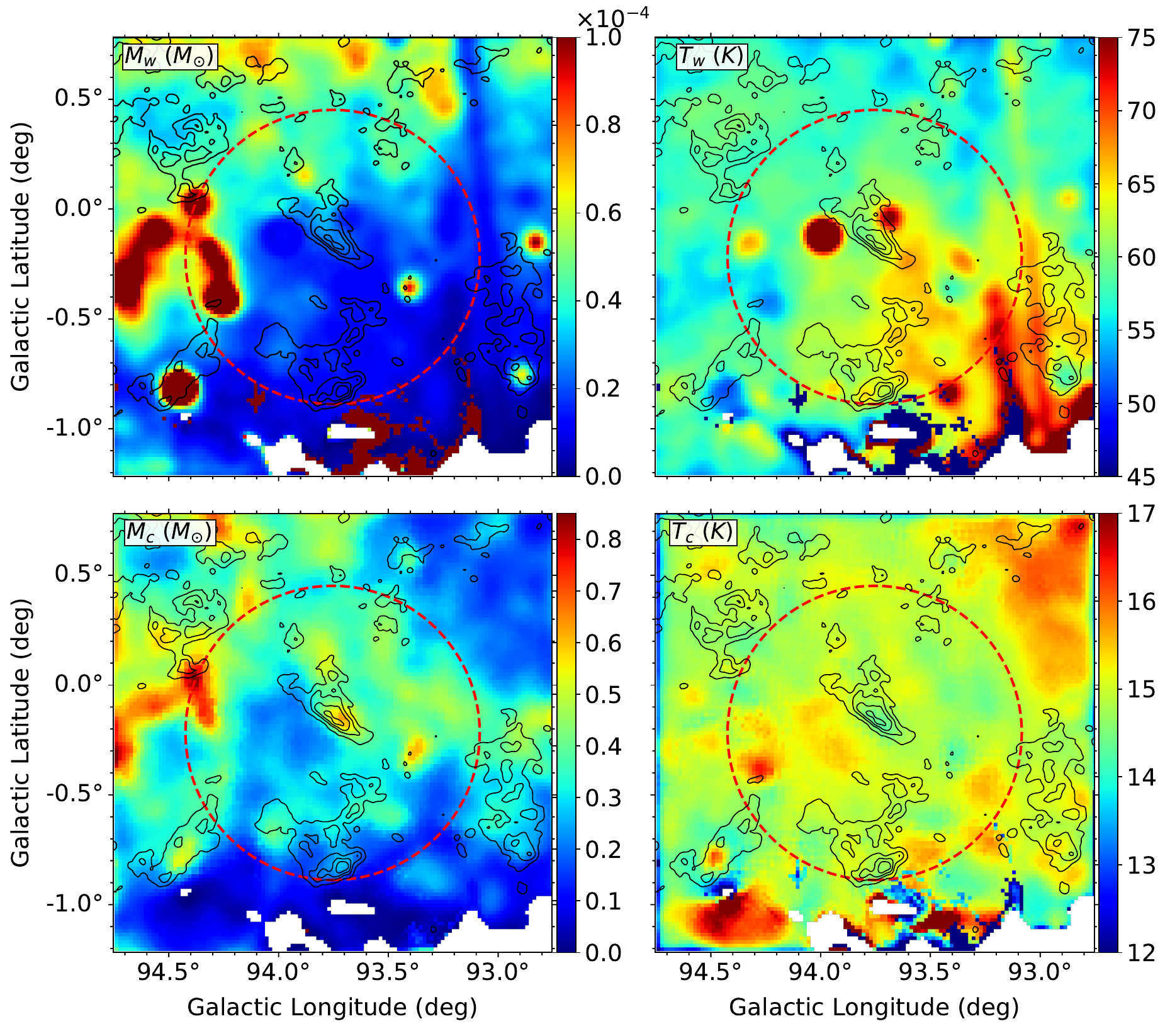}
\caption{Maps of the dust mass (in $M_{\odot}\,{\rm pixel}^{-1}$) and temperature (in K) of G93.7$-$0.2 derived from two-component modified-blackbody fits to the mid-IR--to--submillimeter SEDs. $M_{\rm c}$ ($M_{\odot}$) and $T_{\rm c}$ (K) denote the mass and temperature of the cold dust component, while $M_{\rm w}$ ($M_{\odot}$) and $T_{\rm w}$ (K) refer to the warm dust component. All panels are overlaid with contours of the velocity-integrated $^{12}{\rm CO}$ emission at levels of 3.0, 10.0, 17.0, and 24.0\,K\,km\,s$^{-1}$. The red dashed circle with a radius of \(40\farcm2\) marks the photometric aperture.}
\label{fig5}
\end{figure}

\begin{figure}[ht!]
\centering
\includegraphics[width=1.0\textwidth]{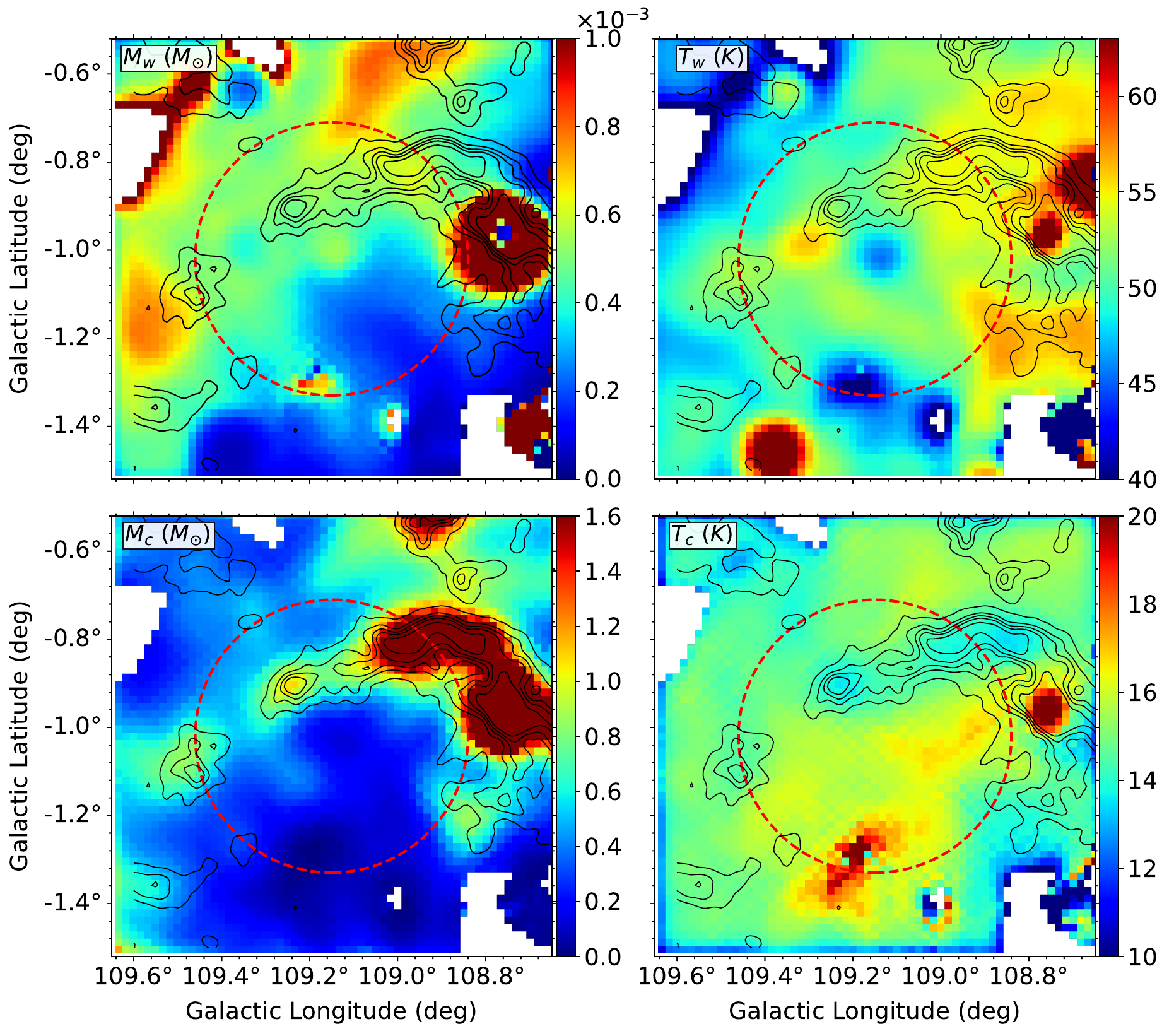}
\caption{Maps of the dust mass (in $M_{\odot}\,{\rm pixel}^{-1}$) and temperature (in K) of G109.1$-$1.0 derived from two-component modified-blackbody fits to the mid-IR--to--submillimeter SEDs. $M_{\rm c}$ ($M_{\odot}$) and $T_{\rm c}$ (K) denote the mass and temperature of the cold dust component, while $M_{\rm w}$ ($M_{\odot}$) and $T_{\rm w}$ (K) refer to the warm dust component. All panels are overlaid with contours of the velocity-integrated $^{12}{\rm CO}$ emission at levels of 3.0, 10.0, 17.0, 24.0, 31.0, and 38.0\,K\,km\,s$^{-1}$. The red dashed circle with a radius of \(18\farcm6\) marks the photometric aperture.}
\label{fig6}
\end{figure}

\begin{figure}[ht!]
\centering
\includegraphics[width=1.0\textwidth]{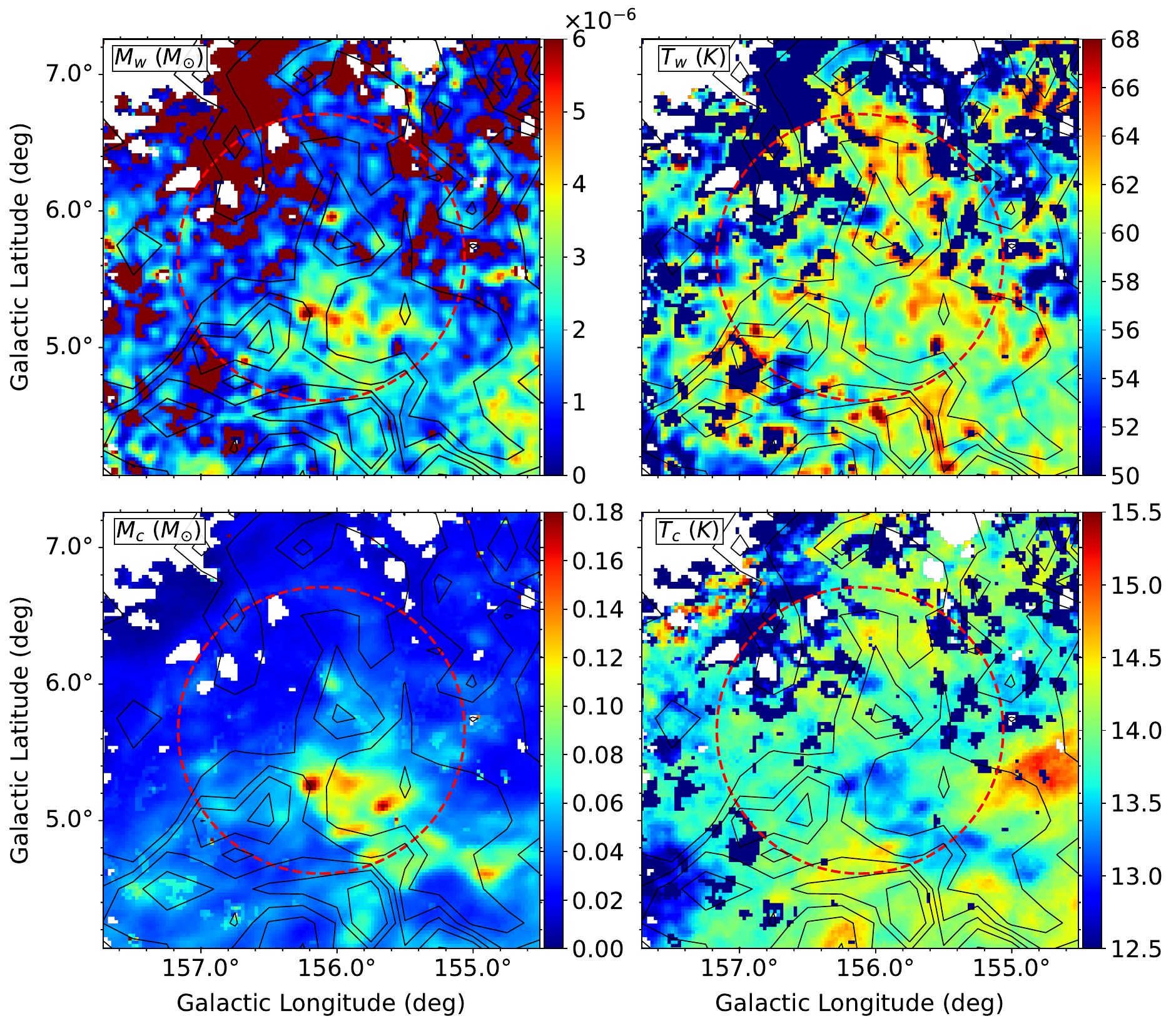}
\caption{Maps of the dust mass (in $M_{\odot}\,{\rm pixel}^{-1}$) and temperature (in K) of G156.2+5.7 derived from two-component modified-blackbody fits to the mid-IR--to--submillimeter SEDs. $M_{\rm c}$ ($M_{\odot}$) and $T_{\rm c}$ (K) denote the mass and temperature of the cold dust component, while $M_{\rm w}$ ($M_{\odot}$) and $T_{\rm w}$ (K) refer to the warm dust component. All panels are overlaid with contours of the velocity-integrated $^{12}{\rm CO}$ emission at levels of 1.0, 2.5, 4.0, 5.5, 7.0, and 8.5\,K\,km\,s$^{-1}$. The red dashed circle with a radius of \(63\arcmin\) marks the photometric aperture.}
\label{fig7}
\end{figure}

\begin{figure}[ht!]
\centering
\includegraphics[width=1.0\textwidth]{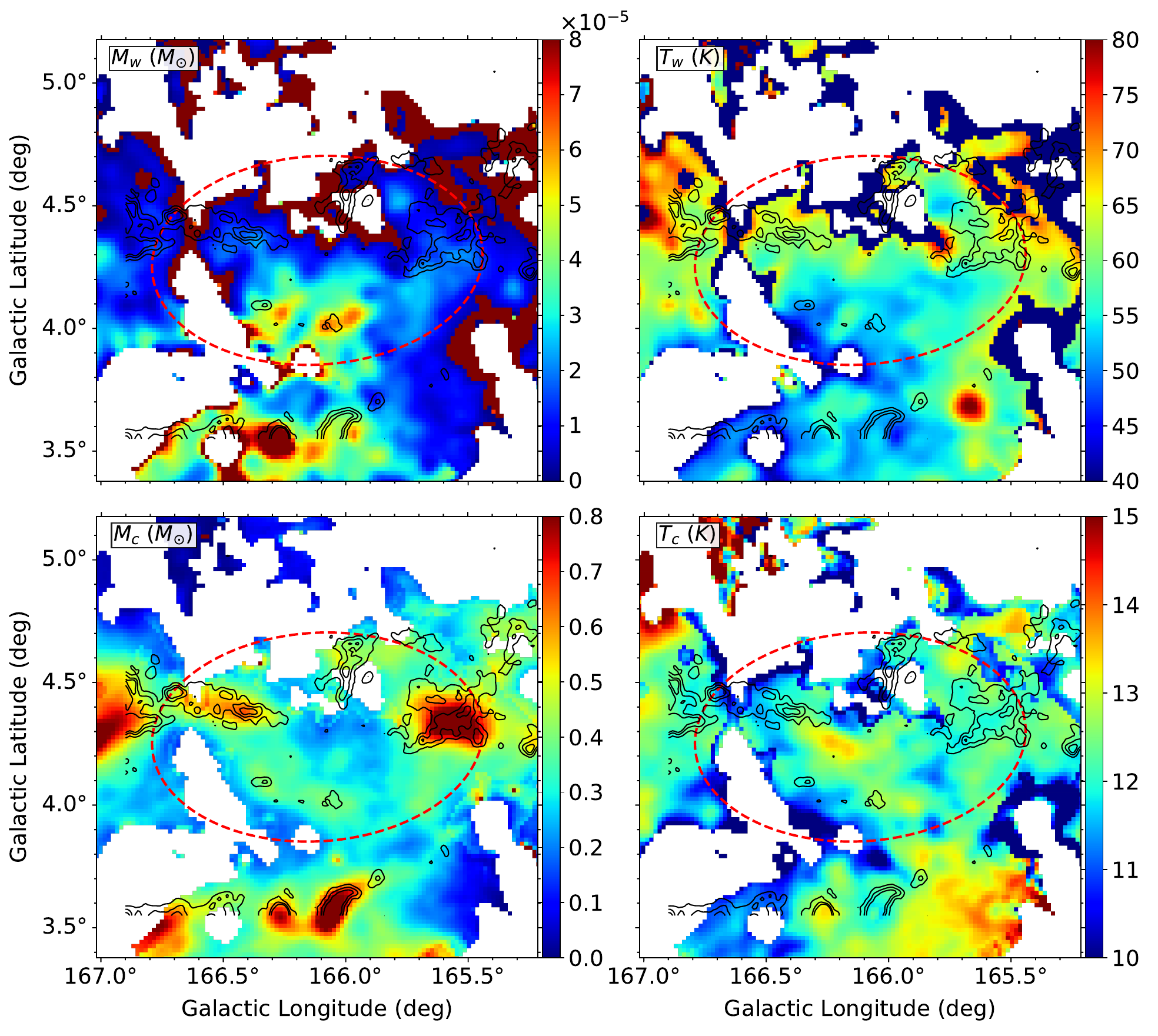}
\caption{Maps of the dust mass (in $M_{\odot}\,{\rm pixel}^{-1}$) and temperature (in K) of G166.0+4.3 derived from two-component modified-blackbody fits to the mid-IR--to--submillimeter SEDs. $M_{\rm c}$ ($M_{\odot}$) and $T_{\rm c}$ (K) denote the mass and temperature of the cold dust component, while $M_{\rm w}$ ($M_{\odot}$) and $T_{\rm w}$ (K) refer to the warm dust component. All panels are overlaid with contours of the velocity-integrated $^{12}{\rm CO}$ emission at levels of 1.5, 3.5, 5.5, and 7.5\,K\,km\,s$^{-1}$. The red dashed ellipse, with semi-major and semi-minor axes of \(40\farcm5\) and \(25\farcm5\) respectively, marks the photometric aperture.}
\label{fig8}
\end{figure}

The spatial distributions of the masses and temperatures of the warm and cold dust components are shown in Figures~\ref{fig5}--\ref{fig8}, with the \(\CO\) velocity-integrated intensity maps of the SNR-associated MCs from Paper~\citetalias{Zhang2026AJ} overlaid for comparison. In all four remnants, the cold dust traces the CO emission more closely than the warm dust does, suggesting that the cold component is predominantly associated with the ambient molecular material.

The detailed dust properties of each remnant are discussed below.

\paragraph{G93.7\(-\)0.2.}
As shown in Figure~\ref{fig5}, the cold dust mass dominates within the molecular clumps surrounding the geometric center of the SNR, while the warm-dust contribution inside the remnant is negligible. The \mbox{H\,{\sc ii}} region to the east of the SNR exhibits enhanced emission in both components, with mass surface densities significantly higher than elsewhere. The cold dust temperature is remarkably uniform across the shell, with a typical value of \(\sim\!15\,\mathrm{K}\), suggesting heating dominated by the ambient interstellar radiation field. In contrast, the warm dust reaches temperatures in excess of \(60\,\mathrm{K}\), peaking in the southern part of the remnant. The cold dust morphology shows a tight spatial correlation with the \(\CO\) emission, tracing the molecular environment, whereas the warm dust emission is enhanced in the \mbox{H\,{\sc ii}} region, indicative of a stronger local radiation field and increased dust heating associated with the ionized gas. In addition, regions adjacent to molecular clumps along the southern boundary of the remnant show simultaneously elevated cold and warm dust temperatures, which may reflect ongoing SNR--MC interaction.

\paragraph{G109.1\(-\)1.0.}
As shown in Figure~\ref{fig6}, the cold dust mass distribution closely follows the CO emission, supporting an origin in the surrounding MCs. The warm dust is predominantly concentrated within the giant MC on the west of the SNR, where the bright source corresponding to the \mbox{H\,{\sc ii}} region Sh2-152 is located \citep{Kothes2012ApJ}. Within the SNR boundary, warm dust with temperatures above \(55\,\mathrm{K}\) is preferentially distributed along the edge of the giant MC; this region has been identified as a shock-compressed interface where the SNR blast wave interacts strongly with dense molecular material \citep{Kothes2002ApJ}. In contrast, cold dust at \(\sim\!14\,\mathrm{K}\) traces both the CO ridge and parts of the more extended cloud. Notably, the CO ridge also harbors a substantial warm dust mass, suggesting that the SNR shock has propagated into this region. The temperature maps further reveal sharp gradients along the SNR southern boundary, likely reflecting complex cloud--shock interactions and inhomogeneities in the ambient density.

\paragraph{G156.2+5.7.}
As shown in Figure~\ref{fig7}, the warm dust mass contributes only a minor fraction of the total dust mass over the entire remnant. The peak mass surface densities of both components lie in the southwest, spatially coincident with the high-extinction dust complex identified by \citet{Gerardy2007MNRAS}. This indicates that a significant fraction of the cold dust along this sightline is likely associated with foreground or background molecular material rather than being intrinsic to the remnant. The cold dust, which is highly correlated with the \(\CO\) emission, exhibits a temperature of approximately \(14\,\mathrm{K}\); although the southwestern region hosts the highest dust mass, its temperature is comparatively low. In contrast, the warm dust in the north of the SNR shows an average temperature of \(\sim\!60\,\mathrm{K}\), and its 
high-temperature distribution correlates well with the CO emission. A plausible interpretation is that this warm dust originates from molecular material heated by the SNR shock. The irregular structures and elevated \(\chi^2/{\rm dof}\) values near the map edges most likely arise from the low signal-to-noise ratio and from convolution edge effects.

\paragraph{G166.0+4.3.}
As shown in Figure~\ref{fig8}, the cold dust mass distribution in G166.0+4.3 correlates very well with the \(\CO\) emission. The warm dust contributes only a very small fraction of the total dust mass and is concentrated mainly in the southern region. The cold dust distribution clearly delineates the ``wing" structure of the SNR, suggesting that this feature is composed predominantly of cold dust. Across most of the 
shell, the cold dust maintains a uniform temperature of \(\sim\!13\,\mathrm{K}\). However, both components show elevated temperatures in and around the CO-emitting regions, with the warm dust reaching an average temperature of \(\sim\!55\,\mathrm{K}\), and the highest temperatures coincide spatially with the strongest CO emission. The warm dust in these regions is most plausibly heated by collisional processes as the SNR shock propagates through the MCs. The extreme temperature anomalies near the shell edges likely arise from the low signal-to-noise ratio in those pixels.

\section{Discussion}\label{sec:discuss}

\subsection{Comparison between Extinction-based and Emission-based Dust Masses}\label{dust_compare}

For each remnant, we have derived two independent estimates of the associated dust mass: one from IR emission photometry (Section~\ref{sec:sed_results}) and the other from the differential extinction across the distance interval of the SNR (Section~\ref{sec:ext_results}). 
A systematic discrepancy is found between the two estimates. We note that distance uncertainties enter both methods in the same way, \(M_{\rm dust}\propto D^{2}\), the adopted distance therefore affects both in the similar way and does not contribute to the discrepancy between them.

A similar discrepancy was reported by \citet{Li2022MNRAS} for SN~2010jl, where the extinction mass was likewise lower than the IR-emission mass. They attributed the difference to the blackbody radius providing only a lower limit on the size of the emitting dust shell. In our sample, an additional and probably important explanation is sightline contamination from unrelated dust. As shown in Paper~\citetalias{Zhang2026AJ}, dense MCs are present in front of G93.7$-$0.2 and G156.2+5.7. Because IR photometry integrates the emission along the entire sightline, it cannot distinguish dust physically associated with the SNR from dust residing in foreground or background clouds, or in diffuse Galactic structures. Part of the emission from these unrelated components may therefore be erroneously attributed to the remnant, resulting in an overestimation of the IR-emission mass. This effect is expected to be especially severe for G93.7\(-\)0.2 and G156.2+5.7, where massive MCs lie along the sightline.

However, the lower cold dust temperature derived for G166.0+4.3 in Table \ref{table4} may suggest that a substantial fraction of dust along the sightline is relatively cold and weakly heated. 
Such cold dust may contribute significantly to extinction but produces weak far-IR emission, which may lead to an underestimate of the dust mass derived from IR emission. This effect may be particularly relevant for G166.0+4.3, which exhibits a blow-out morphology and is thought to evolve in a highly inhomogeneous environment \citep{Arias2019AA,Derlopa2020MNRAS}. The low-density surroundings may reduce collisional heating of dust grains, while foreground or unshocked diffuse molecular material within the extinction distance interval can contribute to the extinction mass without producing strong far-IR emission.
Another possibility is that the dust opacity differs from that of the other SNRs.
If a smaller $\kappa$ is adopted in the calculation, the IR-emission mass would increase, potentially coincide with or even exceed the extinction mass, which will be discussed in the next section.

\subsection{Uncertainties of Dust Masses}\label{dust_U}

The dust mass derived from SED fitting is highly sensitive to the adopted mass absorption coefficient $\kappa$ (Equation~\ref{equ:kapp2}). 
Indeed, it is found that the mass absorption coefficient varies with the dust properties and its environment. For instance, the dust emissivity index $\beta$ is known to anti-correlate with dust temperature \citep{Shetty2009ApJ}.
We calculated the cold dust masses of all SNRs using different values of $\kappa$ summarized in \citet{Li2022ApJ}, and the resulting dust masses are listed in Table~\ref{table5}. 
These comparisons demonstrate that different dust opacity models can even result in an order-of-magnitude difference in the IR-emission mass for the same SNR. 

This dispersion should be regarded as a systematic model uncertainty rather than as a statistical uncertainty. The alternative opacity prescriptions are not intended to define a statistical probability distribution for the dust mass; instead, they provide a test of the dependence of the IR-emission mass on the assumed dust properties and allow us to assess its consistency with the extinction mass.
For each SNR, since the extinction mass is nearly independent of the far-IR dust emissivity, variations in the mass absorption coefficient do not affect the extinction mass. We adopt 500$\mum$ as the reference wavelength for $\kappa$ because this choice is widely used in previous studies and is less sensitive to dust temperature.  
We retain the opacity model of \citet{Draine1984ApJ} as our fiducial prescription because it provides a physically motivated reference for the dust populations considered here.
Accordingly, the uncertainties quoted for the fiducial IR-emission masses represent the statistical uncertainties from the bootstrap analysis, while the range spanned by the alternative opacity prescriptions in Table~\ref{table5} represents an additional systematic uncertainty associated with the dust model. Both sources of uncertainty are considered when comparing the IR-emission and extinction dust masses.
As discussed previously, the $\kappa$ of \citet{Draine1984ApJ} for G166.0+4.3 results in an emission mass lower than the extinction mass. However, this discrepancy would disappear by adopting the opacity model of \citet{Li2001ApJ}. This comparison shows that the difference between the emission and extinction masses is sensitive to the adopted dust opacity. 

Several additional factors may also cause uncertainty in the IR-emission mass. IR cold dust mass is strongly sensitive to the temperature. However, the SNR contains a continuum distribution of grain temperatures, a simplified two-temperature model may yield biased dust mass estimates \citep{Mattsson2015MNRAS}.
Besides, uncertainties in background subtraction and contamination from diffuse Galactic emission, which are particularly difficult to mitigate at the limited angular resolution of the far-IR data, may additionally bias the derived masses upward \citep{Jeong2005MNRAS}.

In addition, the extinction mass measurements are also subject to their own systematic limitations, including the finite distance resolution of the 3D extinction maps, the spatial sampling of background stars, and possible saturation effects in regions of high optical depth \citep{Green2019ApJ,Zhao2020ApJ}. 

Taking these considerations together, we regard the extinction masses as conservative estimates of the dust truly associated with the distance interval of each SNR, whereas the IR-emission masses may include contributions from unrelated dust along the sightline. 

The dust masses derived in this work provide an important basis for future investigations of dust destruction in SNRs. However, quantitative dust destruction efficiency requires additional constraints on the initial dust reservoir, pre-shock ISM density, shock evolution, and dust processing mechanisms. 
In particular, the dust destruction efficiency can depend on the circumstellar structure and the dynamical evolution of the shock \citep{Martinez-Gonzalez2019ApJ}.
A systematic analysis of these factors, combined with a larger sample of SNRs, will be pursued in future studies to better quantify the role of SNR shocks in the destruction and recycling of interstellar dust.

\section{Summary} \label{sec:sum}

This study presents dust mass estimates for four Galactic SNRs---G93.7$-$0.2, G109.1$-$1.0, G156.2+5.7, and G166.0+4.3. We combine two complementary methods: IR emission fitting and optical/near-IR extinction mapping. The extinction method traces the dust column density directly and is independent of the dust temperature, providing a complementary and conservative estimate of the dust mass associated with the SNR distance interval. The IR method, based on multi-band SED fitting to \textit{WISE}, \textit{IRAS}, \textit{AKARI}, and \textit{Planck} photometry, separates warm and cold dust components and provides spatial information on the dust emission, although it may include unrelated dust along the sightline.

For G93.7$-$0.2, G109.1$-$1.0, G156.2+5.7, and G166.0+4.3, the warm dust components have temperatures and masses of \(\sim 74~{\rm K}\) (\(6\times10^{-3}~M_{\odot}\)), \(60~{\rm K}\) (\(4\times10^{-2}~M_{\odot}\)), \(64~{\rm K}\) (\(2\times10^{-3}~M_{\odot}\)), and \(43~{\rm K}\) (\(7\times10^{-2}~M_{\odot}\)) respectively, while the cold dust components have temperatures and masses of \(\sim 14~{\rm K}\) (\(399.8~M_{\odot}\)), \(16~{\rm K}\) (\(119.6~M_{\odot}\)), \(13~{\rm K}\) (\(118.3~M_{\odot}\)), and \(13~{\rm K}\) (\(91.8~M_{\odot}\)).
The extinction masses (\(48.8\)--\(119.2~M_{\odot}\)) are comparable to the IR-emission masses, with the extinction mass values being lower for three remnants but higher for G166.0+4.3.

This discrepancy suggests that the IR-emission masses are likely affected by additional contributions from unrelated foreground or background dust, together with uncertainties associated with SED modeling and dust opacities.
We therefore adopt the extinction mass values as conservative estimates of the dust mass associated with each remnant.

\begin{acknowledgments}
We are grateful to Drs. Shu Wang and Jian Gao for their helpful discussion and suggestion. This work is supported by the National SKA Program of China No. 2025SKA0140100, the National Natural Science Foundation of China (NSFC) project Nos. 12133002 and 12403026. 

\end{acknowledgments}

\vspace{5mm}
\facilities{WISE, AKARI, IRAS, Planck}

\software{Astropy \citep{Astropy2013AA},
          Photutils \citep{Bradley2024zndo},
          APLpy \citep{Robitaille2012ascl.soft},
          dustmaps \citep{Green2018JOSS},
          MontagePy \citep{Good2019ASPC}
          }

\bibliography{sample701}{}

@ARTICLE{Draine1984ApJ,
       author = {{Draine}, B.~T. and {Lee}, H.~M.},
        title = "{Optical Properties of Interstellar Graphite and Silicate Grains}",
      journal = {\apj},
         year = 1984,
        month = oct,
       volume = {285},
        pages = {89},
          doi = {10.1086/162480},
       adsurl = {https://ui.adsabs.harvard.edu/abs/1984ApJ...285...89D}
}

@ARTICLE{Barlow2010AA,
       author = {{Barlow}, M.~J. and {Krause}, O. and {Swinyard}, B.~M. and {Sibthorpe}, B. and {Besel}, M.-A. and {Wesson}, R. and {Ivison}, R.~J. and {Dunne}, L. and {Gear}, W.~K. and {Gomez}, H.~L. and {Hargrave}, P.~C. and {Henning}, Th. and {Leeks}, S.~J. and {Lim}, T.~L. and {Olofsson}, G. and {Polehampton}, E.~T.},
        title = "{A Herschel PACS and SPIRE study of the dust content of the Cassiopeia A supernova remnant}",
      journal = {\aap},
         year = 2010,
        month = jul,
       volume = {518},
          eid = {L138},
        pages = {L138},
          doi = {10.1051/0004-6361/201014585},
archivePrefix = {arXiv},
       eprint = {1005.2688},
 primaryClass = {astro-ph.GA},
       adsurl = {https://ui.adsabs.harvard.edu/abs/2010A&A...518L.138B}
}

@ARTICLE{Purcell1976ApJ,
       author = {{Purcell}, E.~M.},
        title = "{Temperature fluctuations in very small interstellar grains.}",
      journal = {\apj},
         year = 1976,
        month = jun,
       volume = {206},
        pages = {685-690},
          doi = {10.1086/154428},
       adsurl = {https://ui.adsabs.harvard.edu/abs/1976ApJ...206..685P}
}

@ARTICLE{Hildebrand1983QJRAS,
       author = {{Hildebrand}, R.~H.},
        title = "{The determination of cloud masses and dust characteristics from submillimetre thermal emission.}",
      journal = {\qjras},
         year = 1983,
        month = sep,
       volume = {24},
        pages = {267-282},
       adsurl = {https://ui.adsabs.harvard.edu/abs/1983QJRAS..24..267H}
}

@ARTICLE{Dwek1987ApJ,
       author = {{Dwek}, E. and {Dinerstein}, H.~L. and {Gillett}, F.~C. and {Hauser}, M.~G. and {Rice}, W.~L.},
        title = "{Physical Processes and Infrared Emission from the Cassiopeia A Supernova Remnant}",
      journal = {\apj},
         year = 1987,
        month = apr,
       volume = {315},
        pages = {571},
          doi = {10.1086/165160},
       adsurl = {https://ui.adsabs.harvard.edu/abs/1987ApJ...315..571D}
}

@ARTICLE{Gomez2012ApJ,
       author = {{Gomez}, H.~L. and {Krause}, O. and {Barlow}, M.~J. and {Swinyard}, B.~M. and {Owen}, P.~J. and {Clark}, C.~J.~R. and {Matsuura}, M. and {Gomez}, E.~L. and {Rho}, J. and {Besel}, M.-A. and {Bouwman}, J. and {Gear}, W.~K. and {Henning}, Th. and {Ivison}, R.~J. and {Polehampton}, E.~T. and {Sibthorpe}, B.},
        title = "{A Cool Dust Factory in the Crab Nebula: A Herschel Study of the Filaments}",
      journal = {\apj},
         year = 2012,
        month = nov,
       volume = {760},
       number = {1},
          eid = {96},
        pages = {96},
          doi = {10.1088/0004-637X/760/1/96},
archivePrefix = {arXiv},
       eprint = {1209.5677},
 primaryClass = {astro-ph.GA},
       adsurl = {https://ui.adsabs.harvard.edu/abs/2012ApJ...760...96G}
}

@ARTICLE{deLooze2017MNRAS,
       author = {{De Looze}, I. and {Barlow}, M.~J. and {Swinyard}, B.~M. and {Rho}, J. and {Gomez}, H.~L. and {Matsuura}, M. and {Wesson}, R.},
        title = "{The dust mass in Cassiopeia A from a spatially resolved Herschel analysis}",
      journal = {\mnras},
         year = 2017,
        month = mar,
       volume = {465},
       number = {3},
        pages = {3309-3342},
          doi = {10.1093/mnras/stw2837},
archivePrefix = {arXiv},
       eprint = {1611.00774},
 primaryClass = {astro-ph.GA},
       adsurl = {https://ui.adsabs.harvard.edu/abs/2017MNRAS.465.3309D}
}

@ARTICLE{Nozawa2007ApJ,
       author = {{Nozawa}, Takaya and {Kozasa}, Takashi and {Habe}, Asao and {Dwek}, Eli and {Umeda}, Hideyuki and {Tominaga}, Nozomu and {Maeda}, Keiichi and {Nomoto}, Ken'ichi},
        title = "{Evolution of Dust in Primordial Supernova Remnants: Can Dust Grains Formed in the Ejecta Survive and Be Injected into the Early Interstellar Medium?}",
      journal = {\apj},
         year = 2007,
        month = sep,
       volume = {666},
       number = {2},
        pages = {955-966},
          doi = {10.1086/520621},
archivePrefix = {arXiv},
       eprint = {0706.0383},
 primaryClass = {astro-ph},
       adsurl = {https://ui.adsabs.harvard.edu/abs/2007ApJ...666..955N}
}

@ARTICLE{Dwek1986ApJ,
       author = {{Dwek}, E.},
        title = "{Temperature Fluctuations and Infrared Emission from Dust Particles in a Hot Gas}",
      journal = {\apj},
         year = 1986,
        month = mar,
       volume = {302},
        pages = {363},
          doi = {10.1086/163995},
       adsurl = {https://ui.adsabs.harvard.edu/abs/1986ApJ...302..363D}
}

@ARTICLE{Slavin2015ApJ,
       author = {{Slavin}, Jonathan D. and {Dwek}, Eli and {Jones}, Anthony P.},
        title = "{Destruction of Interstellar Dust in Evolving Supernova Remnant Shock Waves}",
      journal = {\apj},
         year = 2015,
        month = apr,
       volume = {803},
       number = {1},
          eid = {7},
        pages = {7},
          doi = {10.1088/0004-637X/803/1/7},
archivePrefix = {arXiv},
       eprint = {1502.00929},
 primaryClass = {astro-ph.GA},
       adsurl = {https://ui.adsabs.harvard.edu/abs/2015ApJ...803....7S}
}

@ARTICLE{Jones1994ApJ,
       author = {{Jones}, A.~P. and {Tielens}, A.~G.~G.~M. and {Hollenbach}, D.~J. and {McKee}, C.~F.},
        title = "{Grain Destruction in Shocks in the Interstellar Medium}",
      journal = {\apj},
         year = 1994,
        month = oct,
       volume = {433},
        pages = {797},
          doi = {10.1086/174689},
       adsurl = {https://ui.adsabs.harvard.edu/abs/1994ApJ...433..797J}
}

@ARTICLE{Mattsson2015MNRAS,
       author = {{Mattsson}, Lars and {Gomez}, Haley L. and {Andersen}, Anja C. and {Matsuura}, Mikako},
        title = "{From flux to dust mass: Does the grain-temperature distribution matter for estimates of cold dust masses in supernova remnants?}",
      journal = {\mnras},
         year = 2015,
        month = jun,
       volume = {449},
       number = {4},
        pages = {4079-4090},
          doi = {10.1093/mnras/stv487},
archivePrefix = {arXiv},
       eprint = {1504.02664},
 primaryClass = {astro-ph.SR},
       adsurl = {https://ui.adsabs.harvard.edu/abs/2015MNRAS.449.4079M}
}

@ARTICLE{Guhathakurta1989ApJ,
       author = {{Guhathakurta}, P. and {Draine}, B.~T.},
        title = "{Temperature Fluctuations in the Interstellar Grains. I. Computational Method and Sublimation of Small Grains}",
      journal = {\apj},
         year = 1989,
        month = oct,
       volume = {345},
        pages = {230},
          doi = {10.1086/167899},
       adsurl = {https://ui.adsabs.harvard.edu/abs/1989ApJ...345..230G}
}

@ARTICLE{Dwek1996ApJ,
       author = {{Dwek}, Eli and {Foster}, Scott M. and {Vancura}, Olaf},
        title = "{Cooling, Sputtering, and Infrared Emission from Dust Grains in Fast Nonradiative Shocks}",
      journal = {\apj},
         year = 1996,
        month = jan,
       volume = {457},
        pages = {244},
          doi = {10.1086/176725},
       adsurl = {https://ui.adsabs.harvard.edu/abs/1996ApJ...457..244D}
}

@ARTICLE{Li2022MNRAS,
       author = {{Li}, Jun and {Gao}, Jian and {Jiang}, Biwei and {Lin}, Zesen},
        title = "{Dust models for the extinction of Type IIn supernova SN 2010jl}",
      journal = {\mnras},
         year = 2022,
        month = apr,
       volume = {511},
       number = {2},
        pages = {2021-2032},
          doi = {10.1093/mnras/stac220},
archivePrefix = {arXiv},
       eprint = {2201.09298},
 primaryClass = {astro-ph.SR},
       adsurl = {https://ui.adsabs.harvard.edu/abs/2022MNRAS.511.2021L}
}

@ARTICLE{Temim2017ApJ,
       author = {{Temim}, Tea and {Dwek}, Eli and {Arendt}, Richard G. and {Borkowski}, Kazimierz J. and {Reynolds}, Stephen P. and {Slane}, Patrick and {Gelfand}, Joseph D. and {Raymond}, John C.},
        title = "{A Massive Shell of Supernova-formed Dust in SNR G54.1+0.3}",
      journal = {\apj},
         year = 2017,
        month = feb,
       volume = {836},
       number = {1},
          eid = {129},
        pages = {129},
          doi = {10.3847/1538-4357/836/1/129},
archivePrefix = {arXiv},
       eprint = {1701.01117},
 primaryClass = {astro-ph.GA},
       adsurl = {https://ui.adsabs.harvard.edu/abs/2017ApJ...836..129T}
}

@ARTICLE{Zhao2025ApJL,
       author = {{Zhao}, He and {Chen}, Bingqiu and {Li}, Jun},
        title = "{Observational Evidence of Dust Evolution in Supernova Remnants: Size Redistribution toward Larger Grains in the Early Sedov Phase}",
      journal = {\apjl},
         year = 2025,
        month = oct,
       volume = {991},
       number = {2},
          eid = {L36},
        pages = {L36},
          doi = {10.3847/2041-8213/ae06fe},
archivePrefix = {arXiv},
       eprint = {2509.12067},
 primaryClass = {astro-ph.GA},
       adsurl = {https://ui.adsabs.harvard.edu/abs/2025ApJ...991L..36Z}
}

@ARTICLE{Matsuura2011Sci,
       author = {{Matsuura}, M. and {Dwek}, E. and {Meixner}, M. and {Otsuka}, M. and {Babler}, B. and {Barlow}, M.~J. and {Roman-Duval}, J. and {Engelbracht}, C. and {Sandstrom}, K. and {Laki{\'c}evi{\'c}}, M. and {van Loon}, J. Th. and {Sonneborn}, G. and {Clayton}, G.~C. and {Long}, K.~S. and {Lundqvist}, P. and {Nozawa}, T. and {Gordon}, K.~D. and {Hony}, S. and {Panuzzo}, P. and {Okumura}, K. and {Misselt}, K.~A. and {Montiel}, E. and {Sauvage}, M.},
        title = "{Herschel Detects a Massive Dust Reservoir in Supernova 1987A}",
      journal = {Science},
         year = 2011,
        month = sep,
       volume = {333},
       number = {6047},
        pages = {1258},
          doi = {10.1126/science.1205983},
archivePrefix = {arXiv},
       eprint = {1107.1477},
 primaryClass = {astro-ph.SR},
       adsurl = {https://ui.adsabs.harvard.edu/abs/2011Sci...333.1258M}
}

@ARTICLE{Derlopa2020MNRAS,
       author = {{Derlopa}, S. and {Boumis}, P. and {Chiotellis}, A. and {Steffen}, W. and {Akras}, S.},
        title = "{First 3D morpho-kinematic model of supernova remnants. The case of VRO 42.05.01 (G166.0+4.3)}",
      journal = {\mnras},
         year = 2020,
        month = dec,
       volume = {499},
       number = {4},
        pages = {5410-5415},
          doi = {10.1093/mnras/staa2336},
archivePrefix = {arXiv},
       eprint = {2008.04958},
 primaryClass = {astro-ph.HE},
       adsurl = {https://ui.adsabs.harvard.edu/abs/2020MNRAS.499.5410D}
}

@ARTICLE{Gomez2012MNRAS,
       author = {{Gomez}, H.~L. and {Clark}, C.~J.~R. and {Nozawa}, T. and {Krause}, O. and {Gomez}, E.~L. and {Matsuura}, M. and {Barlow}, M.~J. and {Besel}, M.-A. and {Dunne}, L. and {Gear}, W.~K. and {Hargrave}, P. and {Henning}, Th. and {Ivison}, R.~J. and {Sibthorpe}, B. and {Swinyard}, B.~M. and {Wesson}, R.},
        title = "{Dust in historical Galactic Type Ia supernova remnants with Herschel}",
      journal = {\mnras},
         year = 2012,
        month = mar,
       volume = {420},
       number = {4},
        pages = {3557-3573},
          doi = {10.1111/j.1365-2966.2011.20272.x},
archivePrefix = {arXiv},
       eprint = {1111.6627},
 primaryClass = {astro-ph.GA},
       adsurl = {https://ui.adsabs.harvard.edu/abs/2012MNRAS.420.3557G}
}

@ARTICLE{Chawner2019MNRAS,
       author = {{Chawner}, H. and {Marsh}, K. and {Matsuura}, M. and {Gomez}, H.~L. and {Cigan}, P. and {De Looze}, I. and {Barlow}, M.~J. and {Dunne}, L. and {Noriega-Crespo}, A. and {Rho}, J.},
        title = "{A catalogue of Galactic supernova remnants in the far-infrared: revealing ejecta dust in pulsar wind nebulae}",
      journal = {\mnras},
         year = 2019,
        month = feb,
       volume = {483},
       number = {1},
        pages = {70-118},
          doi = {10.1093/mnras/sty2942},
archivePrefix = {arXiv},
       eprint = {1811.00034},
 primaryClass = {astro-ph.GA},
       adsurl = {https://ui.adsabs.harvard.edu/abs/2019MNRAS.483...70C}
}

@ARTICLE{Lakicevic2015ApJ,
       author = {{Laki{\'c}evi{\'c}}, Ma{\v{s}}a and {van Loon}, Jacco Th. and {Meixner}, Margaret and {Gordon}, Karl and {Bot}, Caroline and {Roman-Duval}, Julia and {Babler}, Brian and {Bolatto}, Alberto and {Engelbracht}, Chad and {Filipovi{\'c}}, Miroslav and {Hony}, Sacha and {Indebetouw}, Remy and {Misselt}, Karl and {Montiel}, Edward and {Okumura}, K. and {Panuzzo}, Pasquale and {Patat}, Ferdinando and {Sauvage}, Marc and {Seale}, Jonathan and {Sonneborn}, George and {Temim}, Tea and {Uro{\v{s}}evi{\'c}}, Dejan and {Zanardo}, Giovanna},
        title = "{The Influence of Supernova Remnants on the Interstellar Medium in the Large Magellanic Cloud Seen at 20-600 {\ensuremath{\mu}}m Wavelengths}",
      journal = {\apj},
         year = 2015,
        month = jan,
       volume = {799},
       number = {1},
          eid = {50},
        pages = {50},
          doi = {10.1088/0004-637X/799/1/50},
archivePrefix = {arXiv},
       eprint = {1410.5709},
 primaryClass = {astro-ph.GA},
       adsurl = {https://ui.adsabs.harvard.edu/abs/2015ApJ...799...50L}
}

@ARTICLE{Chen2017MNRAS,
       author = {{Chen}, B.-Q. and {Liu}, X.-W. and {Ren}, J.-J. and {Yuan}, H.-B. and {Huang}, Y. and {Yu}, B. and {Xiang}, M.-S. and {Wang}, C. and {Tian}, Z.-J. and {Zhang}, H.-W.},
        title = "{Mapping the three-dimensional dust extinction towards the supernova remnant S147 - the S147 dust cloud}",
      journal = {\mnras},
         year = 2017,
        month = dec,
       volume = {472},
       number = {4},
        pages = {3924-3935},
          doi = {10.1093/mnras/stx2287},
archivePrefix = {arXiv},
       eprint = {1709.01065},
 primaryClass = {astro-ph.GA},
       adsurl = {https://ui.adsabs.harvard.edu/abs/2017MNRAS.472.3924C}
}

@ARTICLE{Katsuda2016ApJ,
       author = {{Katsuda}, Satoru and {Tanaka}, Masaomi and {Morokuma}, Tomoki and {Fesen}, Robert and {Milisavljevic}, Dan},
        title = "{Constraining the Age and Distance of the Galactic Supernova Remnant G156.2+5.7 by H{\ensuremath{\alpha}} Expansion Measurements}",
      journal = {\apj},
         year = 2016,
        month = aug,
       volume = {826},
       number = {2},
          eid = {108},
        pages = {108},
          doi = {10.3847/0004-637X/826/2/108},
archivePrefix = {arXiv},
       eprint = {1605.07271},
 primaryClass = {astro-ph.HE},
       adsurl = {https://ui.adsabs.harvard.edu/abs/2016ApJ...826..108K}
}

@ARTICLE{Zhang2026AJ,
       author = {{Zhang}, Zhe and {Li}, Jun and {Jiang}, Biwei and {Zhao}, He and {Liu}, Fupeng},
        title = "{The Extinction Distance of Supernova Remnants in Combination with the CO Line Measurements}",
      journal = {\aj},
         year = 2026,
        month = may,
       volume = {171},
       number = {5},
          eid = {307},
        pages = {307},
          doi = {10.3847/1538-3881/ae5634},
       adsurl = {https://ui.adsabs.harvard.edu/abs/2026AJ....171..307Z}
}

@ARTICLE{Jeong2005MNRAS,
       author = {{Jeong}, Woong-Seob and {Mok Lee}, Hyung and {Pak}, Soojong and {Nakagawa}, Takao and {Minn Kwon}, Suk and {Pearson}, Chris P. and {White}, Glenn J.},
        title = "{Far-infrared detection limits - I. Sky confusion due to Galactic cirrus}",
      journal = {\mnras},
         year = 2005,
        month = feb,
       volume = {357},
       number = {2},
        pages = {535-547},
          doi = {10.1111/j.1365-2966.2005.08627.x},
archivePrefix = {arXiv},
       eprint = {astro-ph/0411431},
 primaryClass = {astro-ph},
       adsurl = {https://ui.adsabs.harvard.edu/abs/2005MNRAS.357..535J}
}

@software{Bradley2024zndo,
       author = {{Bradley}, Larry and {Sip{\H{o}}cz}, Brigitta and {Robitaille}, Thomas and {Tollerud}, Erik and {Vin{\'\i}cius}, Z{\'e} and {Deil}, Christoph and {Barbary}, Kyle and {Wilson}, Tom J and {Busko}, Ivo and {Donath}, Axel and {G{\"u}nther}, Hans Moritz and {Cara}, Mihai and {Lim}, P.~L. and {Me{\ss}linger}, Sebastian and {Conseil}, Simon and {Burnett}, Zach and {Bostroem}, Azalee and {Droettboom}, Michael and {Bray}, E.~M. and {Andersen Bratholm}, Lars and {Ginsburg}, Adam and {Jamieson}, William and {Barentsen}, Geert and {Craig}, Matt and {Morris}, Brett M. and {Perrin}, Marshall and {Rathi}, Shivangee and {Pascual}, Sergio and {Georgiev}, Iskren Y.},
        title = "{astropy/photutils: 2.0.2}",
         year = 2024,
        month = oct,
          eid = {10.5281/zenodo.13989456},
          doi = {10.5281/zenodo.13989456},
      version = {2.0.2},
    publisher = {Zenodo},
       adsurl = {https://ui.adsabs.harvard.edu/abs/2024zndo..13989456B}
}

@ARTICLE{Burrows1994ApJ,
       author = {{Burrows}, David N. and {Guo}, Zhiyu},
        title = "{ROSAT Observations of VRO 42.05.01}",
      journal = {\apjl},
         year = 1994,
        month = jan,
       volume = {421},
        pages = {L19},
          doi = {10.1086/187177},
       adsurl = {https://ui.adsabs.harvard.edu/abs/1994ApJ...421L..19B}
}

@ARTICLE{Xiao2023ApJ,
       author = {{Xiao}, Li and {Zhu}, Ming and {Sun}, Xiao-Hui and {Jiang}, Peng and {Sun}, Chun},
        title = "{FAST Polarization Mapping of the Supernova Remnant VRO 42.05.01}",
      journal = {\apj},
         year = 2023,
        month = aug,
       volume = {952},
       number = {2},
          eid = {94},
        pages = {94},
          doi = {10.3847/1538-4357/acdb6a},
       adsurl = {https://ui.adsabs.harvard.edu/abs/2023ApJ...952...94X}
}

@ARTICLE{Wright2010AJ,
       author = {{Wright}, Edward L. and {Eisenhardt}, Peter R.~M. and {Mainzer}, Amy K. and {Ressler}, Michael E. and {Cutri}, Roc M. and {Jarrett}, Thomas and {Kirkpatrick}, J. Davy and {Padgett}, Deborah and {McMillan}, Robert S. and {Skrutskie}, Michael and {Stanford}, S.~A. and {Cohen}, Martin and {Walker}, Russell G. and {Mather}, John C. and {Leisawitz}, David and {Gautier}, Thomas N., III and {McLean}, Ian and {Benford}, Dominic and {Lonsdale}, Carol J. and {Blain}, Andrew and {Mendez}, Bryan and {Irace}, William R. and {Duval}, Valerie and {Liu}, Fengchuan and {Royer}, Don and {Heinrichsen}, Ingolf and {Howard}, Joan and {Shannon}, Mark and {Kendall}, Martha and {Walsh}, Amy L. and {Larsen}, Mark and {Cardon}, Joel G. and {Schick}, Scott and {Schwalm}, Mark and {Abid}, Mohamed and {Fabinsky}, Beth and {Naes}, Larry and {Tsai}, Chao-Wei},
        title = "{The Wide-field Infrared Survey Explorer (WISE): Mission Description and Initial On-orbit Performance}",
      journal = {\aj},
         year = 2010,
        month = dec,
       volume = {140},
       number = {6},
        pages = {1868-1881},
          doi = {10.1088/0004-6256/140/6/1868},
archivePrefix = {arXiv},
       eprint = {1008.0031},
 primaryClass = {astro-ph.IM},
       adsurl = {https://ui.adsabs.harvard.edu/abs/2010AJ....140.1868W}
}

@ARTICLE{Neugebauer1984ApJ,
       author = {{Neugebauer}, G. and {Habing}, H.~J. and {van Duinen}, R. and {Aumann}, H.~H. and {Baud}, B. and {Beichman}, C.~A. and {Beintema}, D.~A. and {Boggess}, N. and {Clegg}, P.~E. and {de Jong}, T. and {Emerson}, J.~P. and {Gautier}, T.~N. and {Gillett}, F.~C. and {Harris}, S. and {Hauser}, M.~G. and {Houck}, J.~R. and {Jennings}, R.~E. and {Low}, F.~J. and {Marsden}, P.~L. and {Miley}, G. and {Olnon}, F.~M. and {Pottasch}, S.~R. and {Raimond}, E. and {Rowan-Robinson}, M. and {Soifer}, B.~T. and {Walker}, R.~G. and {Wesselius}, P.~R. and {Young}, E.},
        title = "{The Infrared Astronomical Satellite (IRAS) mission.}",
      journal = {\apjl},
         year = 1984,
        month = mar,
       volume = {278},
        pages = {L1-L6},
          doi = {10.1086/184209},
       adsurl = {https://ui.adsabs.harvard.edu/abs/1984ApJ...278L...1N}
}

@ARTICLE{Sasaki2004ApJ,
       author = {{Sasaki}, Manami and {Plucinsky}, Paul P. and {Gaetz}, Terrance J. and {Smith}, Randall K. and {Edgar}, Richard J. and {Slane}, Patrick O.},
        title = "{XMM-Newton Observations of the Galactic Supernova Remnant CTB 109 (G109.1-1.0)}",
      journal = {\apj},
         year = 2004,
        month = dec,
       volume = {617},
       number = {1},
        pages = {322-338},
          doi = {10.1086/425353},
archivePrefix = {arXiv},
       eprint = {astro-ph/0408290},
 primaryClass = {astro-ph},
       adsurl = {https://ui.adsabs.harvard.edu/abs/2004ApJ...617..322S}
}

@ARTICLE{Sasaki2013AA,
       author = {{Sasaki}, M. and {Plucinsky}, P.~P. and {Gaetz}, T.~J. and {Bocchino}, F.},
        title = "{Chandra observation of the Galactic supernova remnant CTB 109 (G109.1-1.0)}",
      journal = {\aap},
         year = 2013,
        month = apr,
       volume = {552},
          eid = {A45},
        pages = {A45},
          doi = {10.1051/0004-6361/201220836},
archivePrefix = {arXiv},
       eprint = {1302.2459},
 primaryClass = {astro-ph.HE},
       adsurl = {https://ui.adsabs.harvard.edu/abs/2013A&A...552A..45S}
}

@ARTICLE{Uyaniker2002ApJ,
       author = {{Uyaniker}, B{\"u}lent and {Kothes}, Roland and {Brunt}, Christopher M.},
        title = "{The Supernova Remnant CTB 104A: Magnetic Field Structure and Interaction with the Environment}",
      journal = {\apj},
         year = 2002,
        month = feb,
       volume = {565},
       number = {2},
        pages = {1022-1034},
          doi = {10.1086/324782},
archivePrefix = {arXiv},
       eprint = {astro-ph/0110001},
 primaryClass = {astro-ph},
       adsurl = {https://ui.adsabs.harvard.edu/abs/2002ApJ...565.1022U}
}

@ARTICLE{Doi2015PASJ,
       author = {{Doi}, Yasuo and {Takita}, Satoshi and {Ootsubo}, Takafumi and {Arimatsu}, Ko and {Tanaka}, Masahiro and {Kitamura}, Yoshimi and {Kawada}, Mitsunobu and {Matsuura}, Shuji and {Nakagawa}, Takao and {Morishima}, Takahiro and {Hattori}, Makoto and {Komugi}, Shinya and {White}, Glenn J. and {Ikeda}, Norio and {Kato}, Daisuke and {Chinone}, Yuji and {Etxaluze}, Mireya and {Cypriano}, Elysandra F.},
        title = "{The AKARI far-infrared all-sky survey maps}",
      journal = {\pasj},
         year = 2015,
        month = jun,
       volume = {67},
       number = {3},
          eid = {50},
        pages = {50},
          doi = {10.1093/pasj/psv022},
archivePrefix = {arXiv},
       eprint = {1503.06421},
 primaryClass = {astro-ph.GA},
       adsurl = {https://ui.adsabs.harvard.edu/abs/2015PASJ...67...50D}
}

@ARTICLE{Planck2016AA,
       author = {{Planck Collaboration} and {Arnaud}, M. and {Ashdown}, M. and {Atrio-Barandela}, F. and {Aumont}, J. and {Baccigalupi}, C. and {Banday}, A.~J. and {Barreiro}, R.~B. and {Battaner}, E. and {Benabed}, K. and {Benoit-L{\'e}vy}, A. and {Bernard}, J. -P. and {Bersanelli}, M. and {Bielewicz}, P. and {Bobin}, J. and {Bond}, J.~R. and {Borrill}, J. and {Bouchet}, F.~R. and {Brogan}, C.~L. and {Burigana}, C. and {Cardoso}, J. -F. and {Catalano}, A. and {Chamballu}, A. and {Chiang}, H.~C. and {Christensen}, P.~R. and {Colombi}, S. and {Colombo}, L.~P.~L. and {Crill}, B.~P. and {Curto}, A. and {Cuttaia}, F. and {Davies}, R.~D. and {Davis}, R.~J. and {de Bernardis}, P. and {de Rosa}, A. and {de Zotti}, G. and {Delabrouille}, J. and {D{\'e}sert}, F. -X. and {Dickinson}, C. and {Diego}, J.~M. and {Donzelli}, S. and {Dor{\'e}}, O. and {Dupac}, X. and {En{\ss}lin}, T.~A. and {Eriksen}, H.~K. and {Finelli}, F. and {Forni}, O. and {Frailis}, M. and {Fraisse}, A.~A. and {Franceschi}, E. and {Galeotta}, S. and {Ganga}, K. and {Giard}, M. and {Giraud-H{\'e}raud}, Y. and {Gonz{\'a}lez-Nuevo}, J. and {G{\'o}rski}, K.~M. and {Gregorio}, A. and {Gruppuso}, A. and {Hansen}, F.~K. and {Harrison}, D.~L. and {Hern{\'a}ndez-Monteagudo}, C. and {Herranz}, D. and {Hildebrandt}, S.~R. and {Hobson}, M. and {Holmes}, W.~A. and {Huffenberger}, K.~M. and {Jaffe}, A.~H. and {Jaffe}, T.~R. and {Keih{\"a}nen}, E. and {Keskitalo}, R. and {Kisner}, T.~S. and {Kneissl}, R. and {Knoche}, J. and {Kunz}, M. and {Kurki-Suonio}, H. and {L{\"a}hteenm{\"a}ki}, A. and {Lamarre}, J. -M. and {Lasenby}, A. and {Lawrence}, C.~R. and {Leonardi}, R. and {Liguori}, M. and {Lilje}, P.~B. and {Linden-V{\o}rnle}, M. and {L{\'o}pez-Caniego}, M. and {Lubin}, P.~M. and {Maino}, D. and {Maris}, M. and {Marshall}, D.~J. and {Martin}, P.~G. and {Mart{\'\i}nez-Gonz{\'a}lez}, E. and {Masi}, S. and {Matarrese}, S. and {Mazzotta}, P. and {Melchiorri}, A. and {Mendes}, L. and {Mennella}, A. and {Migliaccio}, M. and {Miville-Desch{\^e}nes}, M. -A. and {Moneti}, A. and {Montier}, L. and {Morgante}, G. and {Mortlock}, D. and {Munshi}, D. and {Murphy}, J.~A. and {Naselsky}, P. and {Nati}, F. and {Noviello}, F. and {Novikov}, D. and {Novikov}, I. and {Oppermann}, N. and {Oxborrow}, C.~A. and {Pagano}, L. and {Pajot}, F. and {Paladini}, R. and {Pasian}, F. and {Peel}, M. and {Perdereau}, O. and {Perrotta}, F. and {Piacentini}, F. and {Piat}, M. and {Pietrobon}, D. and {Plaszczynski}, S. and {Pointecouteau}, E. and {Polenta}, G. and {Popa}, L. and {Pratt}, G.~W. and {Puget}, J. -L. and {Rachen}, J.~P. and {Reach}, W.~T. and {Reich}, W. and {Reinecke}, M. and {Remazeilles}, M. and {Renault}, C. and {Rho}, J. and {Ricciardi}, S. and {Riller}, T. and {Ristorcelli}, I. and {Rocha}, G. and {Rosset}, C. and {Roudier}, G. and {Rusholme}, B. and {Sandri}, M. and {Savini}, G. and {Scott}, D. and {Stolyarov}, V. and {Sutton}, D. and {Suur-Uski}, A. -S. and {Sygnet}, J. -F. and {Tauber}, J.~A. and {Terenzi}, L. and {Toffolatti}, L. and {Tomasi}, M. and {Tristram}, M. and {Tucci}, M. and {Umana}, G. and {Valenziano}, L. and {Valiviita}, J. and {Van Tent}, B. and {Vielva}, P. and {Villa}, F. and {Wade}, L.~A. and {Yvon}, D. and {Zacchei}, A. and {Zonca}, A.},
        title = "{Planck intermediate results. XXXI. Microwave survey of Galactic supernova remnants}",
      journal = {\aap},
         year = 2016,
        month = feb,
       volume = {586},
          eid = {A134},
        pages = {A134},
          doi = {10.1051/0004-6361/201425022},
archivePrefix = {arXiv},
       eprint = {1409.5746},
 primaryClass = {astro-ph.GA},
       adsurl = {https://ui.adsabs.harvard.edu/abs/2016A&A...586A.134P}
}

@ARTICLE{Uchida2012PASJ,
       author = {{Uchida}, Hiroyuki and {Tsunemi}, Hiroshi and {Katsuda}, Satoru and {Mori}, Koji and {Petre}, Robert and {Yamaguchi}, Hiroya},
        title = "{A Suzaku Study of Ejecta Structure and Origin of Hard X-Ray Emission in the Supernova Remnant G 156.2+5.7}",
      journal = {\pasj},
         year = 2012,
        month = jun,
       volume = {64},
       number = {3},
          eid = {61},
        pages = {61},
          doi = {10.1093/pasj/64.3.61},
archivePrefix = {arXiv},
       eprint = {1205.4188},
 primaryClass = {astro-ph.HE},
       adsurl = {https://ui.adsabs.harvard.edu/abs/2012PASJ...64...61U}
}

@ARTICLE{Green2019ApJ,
       author = {{Green}, Gregory M. and {Schlafly}, Edward and {Zucker}, Catherine and {Speagle}, Joshua S. and {Finkbeiner}, Douglas},
        title = "{A 3D Dust Map Based on Gaia, Pan-STARRS 1, and 2MASS}",
      journal = {\apj},
         year = 2019,
        month = dec,
       volume = {887},
       number = {1},
          eid = {93},
        pages = {93},
          doi = {10.3847/1538-4357/ab5362},
archivePrefix = {arXiv},
       eprint = {1905.02734},
 primaryClass = {astro-ph.GA},
       adsurl = {https://ui.adsabs.harvard.edu/abs/2019ApJ...887...93G}
}

@ARTICLE{Kahane1985AA,
       author = {{Kahane}, C. and {Guilloteau}, S. and {Lucas}, R.},
        title = "{A multiline study of a typical giant molecular cloud : S 147 / S 153.}",
      journal = {\aap},
         year = 1985,
        month = may,
       volume = {146},
        pages = {325-336},
       adsurl = {https://ui.adsabs.harvard.edu/abs/1985A&A...146..325K}
}

@ARTICLE{Katsuda2009PASJ,
       author = {Katsuda{}, Satoru and {Petre}, Robert and {Hwang}, Una and {Yamaguchi}, Hiroya and {Mori}, Koji and {Tsunemi}, Hiroshi},
        title = "{Suzaku Observations of Thermal and Non-Thermal X-Ray Emission from the Middle-Aged Supernova Remnant G156.2+5.7}",
      journal = {\pasj},
         year = 2009,
        month = jan,
       volume = {61},
        pages = {S155},
          doi = {10.1093/pasj/61.sp1.S155},
archivePrefix = {arXiv},
       eprint = {0902.1782},
 primaryClass = {astro-ph.GA},
       adsurl = {https://ui.adsabs.harvard.edu/abs/2009PASJ...61S.155K}
}

@ARTICLE{Leahy2020ApJS,
       author = {{Leahy}, D.~A. and {Ranasinghe}, S. and {Gelowitz}, M.},
        title = "{Evolutionary Models for 43 Galactic Supernova Remnants with Distances and X-Ray Spectra}",
      journal = {\apjs},
         year = 2020,
        month = may,
       volume = {248},
       number = {1},
          eid = {16},
        pages = {16},
          doi = {10.3847/1538-4365/ab8bd9},
archivePrefix = {arXiv},
       eprint = {2003.08998},
 primaryClass = {astro-ph.HE},
       adsurl = {https://ui.adsabs.harvard.edu/abs/2020ApJS..248...16L}
}

@ARTICLE{Arias2019AA,
       author = {{Arias}, M. and {Dom{\v{c}}ek}, V. and {Zhou}, P. and {Vink}, J.},
        title = "{The environment of supernova remnant VRO 42.05.01 as probed with IRAM 30m molecular line observations}",
      journal = {\aap},
         year = 2019,
        month = jul,
       volume = {627},
          eid = {A75},
        pages = {A75},
          doi = {10.1051/0004-6361/201935528},
archivePrefix = {arXiv},
       eprint = {1906.03801},
 primaryClass = {astro-ph.HE},
       adsurl = {https://ui.adsabs.harvard.edu/abs/2019A&A...627A..75A}
}

@ARTICLE{Kothes2012ApJ,
       author = {{Kothes}, R. and {Foster}, T.},
        title = "{A Thorough Investigation of the Distance to the Supernova Remnant CTB109 and Its Pulsar AXP J2301+5852}",
      journal = {\apjl},
         year = 2012,
        month = feb,
       volume = {746},
       number = {1},
          eid = {L4},
        pages = {L4},
          doi = {10.1088/2041-8205/746/1/L4},
       adsurl = {https://ui.adsabs.harvard.edu/abs/2012ApJ...746L...4K}
}

@ARTICLE{Li2001ApJ,
       author = {{Li}, Aigen and {Draine}, B.~T.},
        title = "{Infrared Emission from Interstellar Dust. II. The Diffuse Interstellar Medium}",
      journal = {\apj},
         year = 2001,
        month = jun,
       volume = {554},
       number = {2},
        pages = {778-802},
          doi = {10.1086/323147},
archivePrefix = {arXiv},
       eprint = {astro-ph/0011319},
 primaryClass = {astro-ph},
       adsurl = {https://ui.adsabs.harvard.edu/abs/2001ApJ...554..778L}
}

@ARTICLE{Green2025JApA,
       author = {{Green}, D.~A.},
        title = "{An updated catalogue of 310 Galactic supernova remnants and their statistical properties}",
      journal = {Journal of Astrophysics and Astronomy},
         year = 2025,
        month = jan,
       volume = {46},
       number = {1},
          eid = {14},
        pages = {14},
          doi = {10.1007/s12036-024-10038-4},
archivePrefix = {arXiv},
       eprint = {2411.03367},
 primaryClass = {astro-ph.GA},
       adsurl = {https://ui.adsabs.harvard.edu/abs/2025JApA...46...14G}
}

@ARTICLE{Gregory1980Nature,
       author = {{Gregory}, P.~C. and {Fahlman}, G.~G.},
        title = "{An extraordinary new celestial X-ray source}",
      journal = {\nat},
         year = 1980,
        month = oct,
       volume = {287},
       number = {5785},
        pages = {805-806},
          doi = {10.1038/287805a0},
       adsurl = {https://ui.adsabs.harvard.edu/abs/1980Natur.287..805G}
}

@ARTICLE{Sharpless1959ApJS,
       author = {{Sharpless}, Stewart},
        title = "{A Catalogue of H II Regions.}",
      journal = {\apjs},
         year = 1959,
        month = dec,
       volume = {4},
        pages = {257},
          doi = {10.1086/190049},
       adsurl = {https://ui.adsabs.harvard.edu/abs/1959ApJS....4..257S}
}

@ARTICLE{Wang2020AA,
       author = {{Wang}, Shu and {Zhang}, Chengyu and {Jiang}, Biwei and {Zhao}, He and {Chen}, Bingqiu and {Chen}, Xiaodian and {Gao}, Jian and {Liu}, Jifeng},
        title = "{Distances to the supernova remnants in the inner disk}",
      journal = {\aap},
         year = 2020,
        month = jul,
       volume = {639},
          eid = {A72},
        pages = {A72},
          doi = {10.1051/0004-6361/201936868},
archivePrefix = {arXiv},
       eprint = {2005.08270},
 primaryClass = {astro-ph.GA},
       adsurl = {https://ui.adsabs.harvard.edu/abs/2020A&A...639A..72W}
}

@ARTICLE{Kothes2002ApJ,
       author = {{Kothes}, Roland and {Uyaniker}, B{\"u}lent and {Yar}, Aylin},
        title = "{The Distance to Supernova Remnant CTB 109 Deduced from Its Environment}",
      journal = {\apj},
         year = 2002,
        month = sep,
       volume = {576},
       number = {1},
        pages = {169-175},
          doi = {10.1086/341545},
archivePrefix = {arXiv},
       eprint = {astro-ph/0205034},
 primaryClass = {astro-ph},
       adsurl = {https://ui.adsabs.harvard.edu/abs/2002ApJ...576..169K}
}

@ARTICLE{James2002MNRAS,
       author = {{James}, A. and {Dunne}, L. and {Eales}, S. and {Edmunds}, M.~G.},
        title = "{SCUBA observations of galaxies with metallicity measurements: a new method for determining the relation between submillimetre luminosity and dust mass}",
      journal = {\mnras},
         year = 2002,
        month = sep,
       volume = {335},
       number = {3},
        pages = {753-761},
          doi = {10.1046/j.1365-8711.2002.05660.x},
archivePrefix = {arXiv},
       eprint = {astro-ph/0204519},
 primaryClass = {astro-ph},
       adsurl = {https://ui.adsabs.harvard.edu/abs/2002MNRAS.335..753J}
}

@ARTICLE{Chiang2018ApJ,
       author = {{Chiang}, I-Da and {Sandstrom}, Karin M. and {Chastenet}, J{\'e}r{\'e}my and {Johnson}, L. Clifton and {Leroy}, Adam K. and {Utomo}, Dyas},
        title = "{The Spatially Resolved Dust-to-metals Ratio in M101}",
      journal = {\apj},
         year = 2018,
        month = oct,
       volume = {865},
       number = {2},
          eid = {117},
        pages = {117},
          doi = {10.3847/1538-4357/aadc5f},
archivePrefix = {arXiv},
       eprint = {1808.07164},
 primaryClass = {astro-ph.GA},
       adsurl = {https://ui.adsabs.harvard.edu/abs/2018ApJ...865..117C}
}

@ARTICLE{Priestley2022MNRAS,
       author = {{Priestley}, F.~D. and {Chawner}, H. and {Barlow}, M.~J. and {De Looze}, I. and {Gomez}, H.~L. and {Matsuura}, M.},
        title = "{Properties of shocked dust grains in supernova remnants}",
      journal = {\mnras},
         year = 2022,
        month = oct,
       volume = {516},
       number = {2},
        pages = {2314-2325},
          doi = {10.1093/mnras/stac2408},
archivePrefix = {arXiv},
       eprint = {2208.11137},
 primaryClass = {astro-ph.GA},
       adsurl = {https://ui.adsabs.harvard.edu/abs/2022MNRAS.516.2314P}
}

@ARTICLE{Gerardy2007MNRAS,
       author = {{Gerardy}, Christopher L. and {Fesen}, Robert A.},
        title = "{Discovery of extensive optical emission associated with the X-ray bright, radio faint Galactic SNR G156.2+5.7}",
      journal = {\mnras},
         year = 2007,
        month = apr,
       volume = {376},
       number = {3},
        pages = {929-938},
          doi = {10.1111/j.1365-2966.2007.11494.x},
archivePrefix = {arXiv},
       eprint = {astro-ph/0608522},
 primaryClass = {astro-ph},
       adsurl = {https://ui.adsabs.harvard.edu/abs/2007MNRAS.376..929G}
}

@ARTICLE{Priestley2019MNRAS,
       author = {{Priestley}, F.~D. and {Barlow}, M.~J. and {De Looze}, I.},
        title = "{The mass, location, and heating of the dust in the Cassiopeia A supernova remnant}",
      journal = {\mnras},
         year = 2019,
        month = may,
       volume = {485},
       number = {1},
        pages = {440-451},
          doi = {10.1093/mnras/stz414},
archivePrefix = {arXiv},
       eprint = {1902.01675},
 primaryClass = {astro-ph.GA},
       adsurl = {https://ui.adsabs.harvard.edu/abs/2019MNRAS.485..440P}
}

@ARTICLE{Zhao2020ApJ,
       author = {{Zhao}, He and {Jiang}, Biwei and {Li}, Jun and {Chen}, Bingqiu and {Yu}, Bin and {Wang}, Ye},
        title = "{A Systematic Study of the Dust of Galactic Supernova Remnants. I. The Distance and the Extinction}",
      journal = {\apj},
         year = 2020,
        month = mar,
       volume = {891},
       number = {2},
          eid = {137},
        pages = {137},
          doi = {10.3847/1538-4357/ab75ef},
archivePrefix = {arXiv},
       eprint = {2002.04748},
 primaryClass = {astro-ph.SR},
       adsurl = {https://ui.adsabs.harvard.edu/abs/2020ApJ...891..137Z}
}

@ARTICLE{Li2022ApJ,
       author = {{Li}, Jun and {Jiang}, Biwei and {Zhao}, He},
        title = "{Dust Mass Associated with the Supernova Remnant IC 443 When Emission Meets Extinction}",
      journal = {\apj},
         year = 2022,
        month = mar,
       volume = {927},
       number = {2},
          eid = {226},
        pages = {226},
          doi = {10.3847/1538-4357/ac5325},
archivePrefix = {arXiv},
       eprint = {2202.05174},
 primaryClass = {astro-ph.HE},
       adsurl = {https://ui.adsabs.harvard.edu/abs/2022ApJ...927..226L}
}

@ARTICLE{Zhao2018ApJ,
       author = {{Zhao}, He and {Jiang}, Biwei and {Gao}, Shuang and {Li}, Jun and {Sun}, Mingxu},
        title = "{The Distance to and the Near-infrared Extinction of the Monoceros Supernova Remnant}",
      journal = {\apj},
         year = 2018,
        month = mar,
       volume = {855},
       number = {1},
          eid = {12},
        pages = {12},
          doi = {10.3847/1538-4357/aaacd0},
archivePrefix = {arXiv},
       eprint = {1802.01069},
 primaryClass = {astro-ph.GA},
       adsurl = {https://ui.adsabs.harvard.edu/abs/2018ApJ...855...12Z}
}

@ARTICLE{Ferrand2012AdSpR,
       author = {{Ferrand}, Gilles and {Safi-Harb}, Samar},
        title = "{A census of high-energy observations of Galactic supernova remnants}",
      journal = {Advances in Space Research},
         year = 2012,
        month = may,
       volume = {49},
       number = {9},
        pages = {1313-1319},
          doi = {10.1016/j.asr.2012.02.004},
archivePrefix = {arXiv},
       eprint = {1202.0245},
 primaryClass = {astro-ph.HE},
       adsurl = {https://ui.adsabs.harvard.edu/abs/2012AdSpR..49.1313F}
}

@ARTICLE{Nozawa2013ApJ,
       author = {{Nozawa}, Takaya and {Fukugita}, Masataka},
        title = "{Properties of Dust Grains Probed with Extinction Curves}",
      journal = {\apj},
         year = 2013,
        month = jun,
       volume = {770},
       number = {1},
          eid = {27},
        pages = {27},
          doi = {10.1088/0004-637X/770/1/27},
archivePrefix = {arXiv},
       eprint = {1301.4024},
 primaryClass = {astro-ph.GA},
       adsurl = {https://ui.adsabs.harvard.edu/abs/2013ApJ...770...27N}
}

@ARTICLE{Schlafly2011ApJ,
       author = {{Schlafly}, Edward F. and {Finkbeiner}, Douglas P.},
        title = "{Measuring Reddening with Sloan Digital Sky Survey Stellar Spectra and Recalibrating SFD}",
      journal = {\apj},
         year = 2011,
        month = aug,
       volume = {737},
       number = {2},
          eid = {103},
        pages = {103},
          doi = {10.1088/0004-637X/737/2/103},
archivePrefix = {arXiv},
       eprint = {1012.4804},
 primaryClass = {astro-ph.GA},
       adsurl = {https://ui.adsabs.harvard.edu/abs/2011ApJ...737..103S}
}

@ARTICLE{Wang2022AJ,
       author = {{Wang}, Ye and {Jiang}, Biwei and {Li}, Jun and {Zhao}, He and {Ren}, Yi},
        title = "{The Dust Mass of Supernova Remnants in M31}",
      journal = {\aj},
         year = 2022,
        month = feb,
       volume = {163},
       number = {2},
          eid = {60},
        pages = {60},
          doi = {10.3847/1538-3881/ac35db},
archivePrefix = {arXiv},
       eprint = {2111.02886},
 primaryClass = {astro-ph.GA},
       adsurl = {https://ui.adsabs.harvard.edu/abs/2022AJ....163...60W}
}

@ARTICLE{Shetty2009ApJ,
       author = {{Shetty}, Rahul and {Kauffmann}, Jens and {Schnee}, Scott and {Goodman}, Alyssa A.},
        title = "{The Effect of Noise on the Dust Temperature-Spectral Index Correlation}",
      journal = {\apj},
         year = 2009,
        month = may,
       volume = {696},
       number = {1},
        pages = {676-680},
          doi = {10.1088/0004-637X/696/1/676},
archivePrefix = {arXiv},
       eprint = {0902.0636},
 primaryClass = {astro-ph.GA},
       adsurl = {https://ui.adsabs.harvard.edu/abs/2009ApJ...696..676S}
}

@ARTICLE{Martinez-Gonzalez2019ApJ,
       author = {{Mart{\'\i}nez-Gonz{\'a}lez}, Sergio and {W{\"u}nsch}, Richard and {Silich}, Sergiy and {Tenorio-Tagle}, Guillermo and {Palou{\v{s}}}, Jan and {Ferrara}, Andrea},
        title = "{Supernovae within Pre-existing Wind-blown Bubbles: Dust Injection versus Ambient Dust Destruction}",
      journal = {\apj},
         year = 2019,
        month = dec,
       volume = {887},
       number = {2},
          eid = {198},
        pages = {198},
          doi = {10.3847/1538-4357/ab571b},
archivePrefix = {arXiv},
       eprint = {1911.05079},
 primaryClass = {astro-ph.GA},
       adsurl = {https://ui.adsabs.harvard.edu/abs/2019ApJ...887..198M}
}

@ARTICLE{Clark2016MNRAS,
       author = {{Clark}, Christopher J.~R. and {Schofield}, Simon P. and {Gomez}, Haley L. and {Davies}, Jonathan I.},
        title = "{An empirical determination of the dust mass absorption coefficient, {\ensuremath{\kappa}}$_{d}$, using the Herschel Reference Survey}",
      journal = {\mnras},
         year = 2016,
        month = jun,
       volume = {459},
       number = {2},
        pages = {1646-1658},
          doi = {10.1093/mnras/stw647},
archivePrefix = {arXiv},
       eprint = {1603.04860},
 primaryClass = {astro-ph.GA},
       adsurl = {https://ui.adsabs.harvard.edu/abs/2016MNRAS.459.1646C}
}

@ARTICLE{Astropy2013AA,
       author = {{Astropy Collaboration} and {Robitaille}, Thomas P. and {Tollerud}, Erik J. and {Greenfield}, Perry and {Droettboom}, Michael and {Bray}, Erik and {Aldcroft}, Tom and {Davis}, Matt and {Ginsburg}, Adam and {Price-Whelan}, Adrian M. and {Kerzendorf}, Wolfgang E. and {Conley}, Alexander and {Crighton}, Neil and {Barbary}, Kyle and {Muna}, Demitri and {Ferguson}, Henry and {Grollier}, Fr{\'e}d{\'e}ric and {Parikh}, Madhura M. and {Nair}, Prasanth H. and {Unther}, Hans M. and {Deil}, Christoph and {Woillez}, Julien and {Conseil}, Simon and {Kramer}, Roban and {Turner}, James E.~H. and {Singer}, Leo and {Fox}, Ryan and {Weaver}, Benjamin A. and {Zabalza}, Victor and {Edwards}, Zachary I. and {Azalee Bostroem}, K. and {Burke}, D.~J. and {Casey}, Andrew R. and {Crawford}, Steven M. and {Dencheva}, Nadia and {Ely}, Justin and {Jenness}, Tim and {Labrie}, Kathleen and {Lim}, Pey Lian and {Pierfederici}, Francesco and {Pontzen}, Andrew and {Ptak}, Andy and {Refsdal}, Brian and {Servillat}, Mathieu and {Streicher}, Ole},
        title = "{Astropy: A community Python package for astronomy}",
      journal = {\aap},
         year = 2013,
        month = oct,
       volume = {558},
          eid = {A33},
        pages = {A33},
          doi = {10.1051/0004-6361/201322068},
archivePrefix = {arXiv},
       eprint = {1307.6212},
 primaryClass = {astro-ph.IM},
       adsurl = {https://ui.adsabs.harvard.edu/abs/2013A&A...558A..33A}
}

@ARTICLE{Green2018JOSS,
       author = {{Green}, {Gregory M.}},
        title = "{dustmaps: A Python interface for maps of interstellar dust}",
      journal = {The Journal of Open Source Software},
         year = "2018",
        month = "Jun",
       volume = {3},
       number = {26},
        pages = {695},
          doi = {10.21105/joss.00695},
       adsurl = {https://ui.adsabs.harvard.edu/abs/2018JOSS....3..695G}
}

@software{Robitaille2012ascl.soft,
       author = {{Robitaille}, Thomas and {Bressert}, Eli},
        title = "{APLpy: Astronomical Plotting Library in Python}",
 howpublished = {Astrophysics Source Code Library, record ascl:1208.017},
         year = 2012,
        month = aug,
          eid = {ascl:1208.017},
       adsurl = {https://ui.adsabs.harvard.edu/abs/2012ascl.soft08017R}
}

@INPROCEEDINGS{Good2019ASPC,
       author = {{Good}, John and {Berriman}, G. Bruce},
        title = "{Image Processing in Python with Montage}",
    booktitle = {Astronomical Data Analysis Software and Systems XXVII},
         year = 2019,
       editor = {{Teuben}, Peter J. and {Pound}, Marc W. and {Thomas}, Brian A. and {Warner}, Elizabeth M.},
       series = {Astronomical Society of the Pacific Conference Series},
       volume = {523},
        month = oct,
        pages = {685},
          doi = {10.48550/arXiv.1908.09753},
archivePrefix = {arXiv},
       eprint = {1908.09753},
 primaryClass = {astro-ph.IM},
       adsurl = {https://ui.adsabs.harvard.edu/abs/2019ASPC..523..685G}
}

@ARTICLE{Sanchez2018MNRAS,
       author = {{S{\'a}nchez-Cruces}, M. and {Rosado}, M. and {Fuentes-Carrera}, I. and {Ambrocio-Cruz}, P.},
        title = "{Kinematics of the Galactic Supernova Remnant G109.1-1.0 (CTB 109)}",
      journal = {\mnras},
         year = 2018,
        month = jan,
       volume = {473},
       number = {2},
        pages = {1705-1717},
          doi = {10.1093/mnras/stx2460},
archivePrefix = {arXiv},
       eprint = {1709.07986},
 primaryClass = {astro-ph.GA},
       adsurl = {https://ui.adsabs.harvard.edu/abs/2018MNRAS.473.1705S}
}

@ARTICLE{Gonzalez2025AA,
       author = {{Mart{\'\i}nez-Gonz{\'a}lez}, Sergio},
        title = "{Dusty clump survival in supernova ejecta: Dust-mediated growth versus crushing by the reverse shock}",
      journal = {\aap},
         year = 2025,
        month = oct,
       volume = {702},
          eid = {L6},
        pages = {L6},
          doi = {10.1051/0004-6361/202556389},
archivePrefix = {arXiv},
       eprint = {2509.08887},
 primaryClass = {astro-ph.GA},
       adsurl = {https://ui.adsabs.harvard.edu/abs/2025A&A...702L...6M}
}
\bibliographystyle{aasjournalv7}


\begin{table*}[h]
\renewcommand{\arraystretch}{1.2}
\centering
\caption{Basic Properties of the Four SNRs Analyzed in This Work.}
\label{table1}
\begin{tabular}{c c c c c c c c}
\hline \hline
SNR name &  $R_{\text{SNR}}$ & SN Type & Morphology & Age & Ref & $D$ & $v_{\text{interact}}^{(b)}$\\
   & (pc) &  &  &(kyr) &  &(kpc) & ($\kms$)\\
\hline
G93.7$-$0.2 (CTB104A) &  21.3 & $\cdots$ & shell & $\cdots$ & $\cdots$ &$1.82 \pm 0.13$ & [$-19, -3$] \\
G109.1$-$1.0 (CTB109) &  16.5 & CC$^{(a)}$ & composite & 8.8--14 & 1,2,3 & $3.05 \pm 0.15$ & [$-51, -46$] \\
G156.2+5.7 &  11.0 & CC & shell &15--26 & 4,5,6 &$0.60 \pm 0.15$ & [$-$10, 0] \\
G166.0+4.3 (VRO 42.05.01) &  $40.5 \times 25.5$ & CC$?$ & composite &14.5, 24 & 7,8 & $3.44 \pm 0.23$ & [$-27, -15$] \\
\hline
\end{tabular}
\tablecomments{
\scriptsize 
(a) Core collapse.\\
(b) The velocity components of SNR-MC interaction derived from Paper \citetalias{Zhang2026AJ}.\\
{References:}
(1) \citet{Sasaki2004ApJ}, (2) \citet{Sanchez2018MNRAS}, (3) \citet{Sasaki2013AA},
(4) \citet{Gerardy2007MNRAS}, (5) \citet{Katsuda2009PASJ}, (6) \citet{Katsuda2016ApJ}, 
(7) \citet{Leahy2020ApJS}, (8) \citet{Burrows1994ApJ}\\
}
\end{table*}

\begin{table*}[h]
\centering
\renewcommand{\arraystretch}{1.2}
\centering
\caption{Multiband Infrared Photometry of the Four SNRs from \textit{WISE}, \textit{IRAS}, \textit{AKARI}, and 	\textit{Planck}.}
\label{table2}
\begin{tabular}{c c c c c c c c}
\hline \hline
Band & Resolution & Calibration & Pixel size & \multicolumn{4}{c}{Flux Density (Jy)} \\
\cmidrule(lr){5-8}
(\mum) & (arcsec) & Uncertainty & (arcsec) & G93.7$-$0.2 & G109.1$-$1.0 & G156.2+5.7 & G166.0+4.3\\
\hline
WISE 3.4 & 6.1 & 2.9$\per$ & 1.375 & 167.4$\pm$4.9 & 27.2$\pm$0.8 & 125.4$\pm$3.7 & 45.9$\pm$1.4 \\
WISE 4.6 & 6.4 & 3.4$\per$ & 1.375 & 119.7$\pm$4.1 & 19.4$\pm$0.6 & 78.2$\pm$2.7 & 35.6$\pm$1.2 \\
WISE 12 & 6.5 & 4.6$\per$ & 1.375 & 159.1$\pm$7.3 & 46.7$\pm$2.1 & 108.8$\pm$5.0 & 15.8$\pm$0.7 \\
WISE 22 & 12 & 5.6$\per$ & 1.375 & 212.1$\pm$11.9 & 72.2$\pm$4.1 & 119.9$\pm$6.7 & 1.2$\pm$0.1 \\
\hline
IRAS 12 & 270 & 20$\per$ & 90 & 168.1$\pm$36.9 & 44.3$\pm$9.6 & 117.8$\pm$25.4 & 20.5$\pm$4.2 \\
IRAS 25 & 276 & 20$\per$ & 90 &178.2$\pm$40.3 & 53.9$\pm$11.7 & 134.3$\pm$27.6 & 9.9$\pm$2.6 \\
IRAS 60 & 282 & 20$\per$ & 90 &233.7$\pm$109.2 & 328.9$\pm$75.7 & 378.8$\pm$76.7 & 97.6$\pm$19.8 \\
IRAS 100 & 300 & 20$\per$ & 90 &1965.6$\pm$449.4 & 1060.1$\pm$240.6 & 2891.9$\pm$582.8 & 137.3$\pm$30.8 \\
\hline
AKARI 65 & 63 & 10$\per$ & 15 &749.7$\pm$122.5 & 462.8$\pm$47.5 & 535.8$\pm$55.3 & 74.6$\pm$10.4 \\
AKARI 90 & 78 & 10$\per$ & 15 &962.4$\pm$112.2 & 558.9$\pm$56.9 & 1518.3$\pm$152.0 & 92.5$\pm$9.5 \\
AKARI 140 & 88 & 10$\per$ & 15 &5540.7$\pm$558.3 & 1259.4$\pm$129.2 & 9717.7$\pm$972.3 & 240.2$\pm$25.4 \\
AKARI 160 & 88 & 10$\per$ & 15 &6580.3$\pm$663.0 & 1489.2$\pm$152.9 & 8598.0$\pm$860.8 & 206.9$\pm$24.4 \\
\hline
Planck 350 & 278 & 6.4$\per$ & 60 &3545.7$\pm$252.9 & 585.6$\pm$57.9 & 7593.7$\pm$491.1 & 180.4$\pm$20.3 \\
Planck 550 & 290 & 6.1$\per$ & 60 &1253.6$\pm$86.1 & 109.2$\pm$19.2 & 2842.5$\pm$175.4 & 64.6$\pm$7.5 \\
Planck 850 & 296 & 0.78$\per$ & 60 &371.9$\pm$12.3 & 54.7$\pm$4.5 & 848.9$\pm$10.5 & 24.5$\pm$2.0 \\
Planck 1380 & 301 & 0.16$\per$ & 60 &68.3$\pm$2.9 & 13.7$\pm$1.1 & 207.1$\pm$2.2 & 13.9$\pm$0.6 \\
\hline
\end{tabular}
\end{table*}

\begin{table*}[h!]
\centering
\caption{Aperture Radii and Background Annuli Adopted for the Multiband Photometry of the Four SNRs.}
\label{table3}
\begin{tabular}{c c c c}
\hline \hline
SNR & $R_{\text{aperture}}$ & $R_{\text{inner}}$ & $R_{\text{outer}}$ \\
   & (arcmin) &  (arcmin) & (arcmin) \\
\hline
G93.7$-$0.2 & 40.2 & 48 & 60.6 \\
G109.1$-$1.0 & 18.6 & 21.6 & 27.6 \\
G156.2+5.7 &  63.0 & 76.8 & 94.8 \\
G166.0+4.3 &  $40.5 \times 25.5^{(a)}$ & $47.7 \times 30$ & $57.3 \times 36$  \\
\hline
\end{tabular}
\tablecomments{\scriptsize 
(a) The aperture and background annulus of G166.0+4.3 are elliptical; the two values give the semi- major and semi-minor axes, respectively.
}
\end{table*}

\begin{table*}[h]
\renewcommand{\arraystretch}{1.2}
\centering
\caption{Masses and Temperatures of the Warm and Cold Dust Components of the Four SNRs, Compared with the Dust Masses Derived from Extinction.}
\label{table4}
\begin{tabular}{c c c c c c}
\hline \hline
SNR & $\mw$ & $\tw$ & $\mc$ & $\tc$ & $M^{\text{ext}}_{\text{d}}$ \\
   & ($\msun$) & (K) & ($\msun$) & (K) & ($\msun$) \\
\hline
G93.7$-$0.2 & $0.006_{-0.002}^{+0.003}$ & $74.3_{-2.9}^{+4.2}$ & $399.8_{-15.8}^{+17.5}$ & $14.2_{-0.2}^{+0.2}$ & $108.3^{+17.9}_{-13.5}$ \\
G109.1$-$1.0 & $0.04_{-0.01}^{+0.01}$ & $60.0_{-1.2}^{+1.3}$ & $119.6_{-9.7}^{+10.0}$ & $16.1_{-0.3}^{+0.3}$ & 82.0$^{+13.6}_{-10.2}$ \\
G156.2+5.7 &  $0.002_{-0.0003}^{+0.0003}$ & $63.9_{-0.9}^{+1.0}$ & $118.3_{-2.6}^{+2.7}$ & $13.2_{-0.1}^{+0.1}$ & 48.8$^{+8.1}_{-6.1}$ \\
G166.0+4.3 &  $0.07_{-0.01}^{+0.01}$ & $42.7_{-0.6}^{+0.5}$ & $91.8_{-9.1}^{+9.2}$ & $12.9_{-0.3}^{+0.3}$ & 119.2$^{+19.7}_{-14.8}$ \\
\hline
\end{tabular}
\tablecomments{The dust mass and temperature for each SNR, with the 1\,$\sigma$ as the uncertainty, are obtained by the bootstrap approach. $\mw$ and $\tw$ represent the mass and temperature of warm dust, $\mc$ and $\tc$ represent the mass and temperature of cold dust, $M^{\text{ext}}_{\text{d}}$ represent the mass derived from extinction. \\
}
\end{table*}

\begin{table*}[h]
\renewcommand{\arraystretch}{1.2}
\caption{The Dust Opacity and Dust Mass and Temperature from IR Emission.}
\label{table5}
\begin{tabular}{c c c c c c c c c}
\hline \hline
SNR &  $\mw$ & $\tw$ & $\mc$ & $\tc$ & Literature & $\lambda_0$ & $\kappa_{\lambda_0}$ & $\beta$ \\
   & ($\msun$) & (K) & ($\msun$) & (K) & References & ($\mum$) & ($\cg$) &  \\
\hline
\multirow{4}{*}{G93.7$-$0.2} & $0.005_{-0.002}^{+0.003}$ & $88.5_{-5.4}^{+7.5}$ & $161.2_{-6.8}^{+7.0}$ & $16.5_{-0.2}^{+0.2}$ & a & 125 & 18.9 & 1.53 \\
 & $0.007_{-0.003}^{+0.004}$ & $88.7_{-5.7}^{+8.4}$ & $232.0_{-9.4}^{+9.8}$ & $16.5_{-0.2}^{+0.2}$ & b & 850 & 0.7 & 1.53 \\
  & $0.02_{-0.01}^{+0.01}$ & $88.3_{-5.2}^{+7.3}$ & $731.5_{-29.0}^{+31.0}$ & $16.5_{-0.2}^{+0.2}$ & c & 500 & 0.51 & 1.53 \\
 & $0.006_{-0.002}^{+0.003}$ & $74.3_{-2.9}^{+4.2}$ & $399.8_{-15.8}^{+17.5}$ & $14.2_{-0.2}^{+0.2}$ & d & 500 & 1.45 & 2.04 \\
  & $0.01_{-0.003}^{+0.004}$ & $72.4_{-2.7}^{+3.6}$ & $659.5_{-27.4}^{+28.0}$ & $13.8_{-0.2}^{+0.2}$& e & 500 & 0.95 & 2.13 \\
\hline
\multirow{4}{*}{G109.1$-$1.0} & $0.04_{-0.01}^{+0.01}$ & $67.6_{-1.8}^{+2.2}$ & $52.4_{-3.8}^{+3.8}$ & $20.0_{-0.4}^{+0.3}$ & a & 125 & 18.9 & 1.53 \\
 & $0.06_{-0.01}^{+0.01}$ & $67.6_{-1.6}^{+1.9}$ & $75.1_{-6.0}^{+6.2}$ & $19.1_{-0.4}^{+0.4}$ & b & 850 & 0.7 & 1.53 \\
  & $0.2_{-0.04}^{+0.05}$ & $67.7_{-1.7}^{+2.0}$ & $232.2_{-18.7}^{+18.7}$ & $19.1_{-0.4}^{+0.4}$ & c & 500 & 0.51 & 1.53 \\
 & $0.04_{-0.01}^{+0.01}$ & $60.0_{-1.2}^{+1.3}$ & $119.6_{-9.7}^{+10.0}$ & $16.1_{-0.3}^{+0.3}$ & d & 500 & 1.45 & 2.04 \\
  & $0.06_{-0.01}^{+0.01}$ & $59.1_{-1.0}^{+1.1}$ & $193.8_{-15.7}^{+16.6}$ & $15.7_{-0.3}^{+0.3}$& e & 500 & 0.95 & 2.13 \\
\hline
\multirow{4}{*}{G156.2+5.7} & $0.002_{-0.0004}^{+0.0003}$ & $72.3_{-1.7}^{+1.9}$ & $45.1_{-1.0}^{+1.0}$ & $15.5_{-0.2}^{+0.2}$ & a & 125 & 18.9 & 1.53\\
 & $0.002_{-0.0005}^{+0.0005}$ & $72.3_{-1.7}^{+2.0}$ & $64.1_{-1.4}^{+1.5}$ & $15.5_{-0.2}^{+0.2}$ & b & 850 & 0.7 & 1.53 \\
  & $0.007_{-0.002}^{+0.002}$ & $72.3_{-1.7}^{+1.9}$ & $198.3_{-4.3}^{+4.5}$ & $15.5_{-0.2}^{+0.2}$ & c & 500 & 0.51 & 1.53 \\
 & $0.002_{-0.0003}^{+0.0003}$ & $63.9_{-0.9}^{+1.0}$ & $118.3_{-2.6}^{+2.7}$ & $13.2_{-0.1}^{+0.1}$ & d & 500 & 1.45 & 2.04 \\
  & $0.002_{-0.0003}^{+0.0004}$ & $62.9_{-1.1}^{+1.1}$ & $197.8_{-4.5}^{+4.4}$ & $12.8_{-0.1}^{+0.1}$ & e & 500 & 0.95 & 2.13 \\
\hline
\multirow{4}{*}{G166.0+4.3} & $0.07_{-0.01}^{+0.01}$ & $45.7_{-0.8}^{+0.8}$ & $39.6_{-3.9}^{+4.1}$ & $14.9_{-0.4}^{+0.4}$ & a &  125 & 18.9 & 1.53\\
 & $0.11_{-0.02}^{+0.02}$ & $45.7_{-0.7}^{+0.8}$ & $56.8_{-5.6}^{+6.0}$ & $14.9_{-0.4}^{+0.4}$ & b & 850 & 0.7 & 1.53 \\
  & $0.33_{-0.05}^{+0.05}$ & $45.7_{-0.7}^{+0.8}$ & $176.0_{-16.4}^{+17.0}$ & $14.9_{-0.4}^{+0.4}$ & c & 500 & 0.51 & 1.53 \\
 & $0.07_{-0.01}^{+0.01}$ & $42.7_{-0.6}^{+0.5}$ & $91.8_{-9.1}^{+9.2}$ & $12.9_{-0.3}^{+0.3}$ & d & 500 & 1.45 & 2.04 \\
  & $0.1_{-0.01}^{+0.01}$ & $42.2_{-0.6}^{+0.6}$ & $150.4_{-14.6}^{+15.4}$ & $12.6_{-0.3}^{+0.3}$ & e & 500 & 0.95 & 2.13 \\
\hline

\end{tabular}
\tablecomments{\scriptsize 
\text{References:}
(a) \citet{Hildebrand1983QJRAS}, (b) \citet{James2002MNRAS}, (c) \citet{Clark2016MNRAS}, (d) \citet{Draine1984ApJ}, (e) \citet{Li2001ApJ}}
\end{table*}


\end{CJK*}
\end{document}